\documentclass[pra, aps, superscriptaddress, twocolumn,
               floatfix, footinbib, 10pt,
               longbibliography]{revtex4-1}

\usepackage{hyperref}
\usepackage{natbib}
\usepackage{amsmath}
\usepackage{amssymb}
\usepackage{graphicx}
\usepackage{color}
\usepackage{hepunits}
\usepackage[multidot]{grffile}

\usepackage{tikz}

\newcommand{\cdiff}[1]{\ensuremath{\partial_{#1}}}

\newcommand{\psic}{\ensuremath{\psi_C}} % photonic field
\newcommand{\psix}{\ensuremath{\psi_X}} % excitonic field

\newcommand{\wisb}{\ensuremath{\omega_\mathrm{ISB}}}

\newcommand{\rr}{\mathbf{r}}
\newcommand{\kk}{\mathbf{k}}

\graphicspath{{img/}}

\newcommand{\nothing}[1]{}

\begin{document}
\title{A generalized Gross-Pitaevskii model for intersubband polariton lasing}

\author{Jacopo Nespolo}
\email{e-mail: jacopo.nespolo@unitn.it}

\author{Iacopo Carusotto}
\email{e-mail: iacopo.carusotto@unitn.it}

\affiliation{INO-CNR BEC Center and Dipartimento di Fisica, Universit\`a di 
Trento, I-38123 Povo, Italy
}

\begin{abstract}
We develop a generalized Gross-Pitaevskii approach to the driven-dissipative 
dynamics of intersubband polaritons in patterned planar microcavities where 
the cavity mode is strongly coupled to an intersubband transition in doped 
quantum wells. Substantial differences with respect to the case of interband 
excitonic polaritons are highlighted, in particular the non-Markovian features 
of the radiative decay. The accuracy of the method is validated on the linear 
reflection properties of the cavity, that quantitatively reproduce 
experimental observations. The theoretical framework is then applied in the 
nonlinear regime to study optical parametric oscillation processes for 
intersubband polaritons. Our findings open interesting perspectives in view of 
novel coherent laser sources operating in the mid and far infrared.
\end{abstract}

\date{\today}

\pacs{}
%\keywords{}

\maketitle
%%%%%%%%%%%%%%%%%%%%%%%%%%%%%%%%%%%%%%%%%%%%%%%%%%%%%%%%%%%%%%%%%%%%%%%%%%%%%%%

Planar semiconductor microcavities have emerged as a versatile tool to address
fundamental questions in the physics of light-matter interaction, and realize 
a new generation of optoelectronic devices, with applications that are still 
largely unexplored~\cite{Microcavities}.  In particular, a rich variety of 
optical phenomena are observed when a quantum well element is embedded in the 
cavity layer and an electronic transition is resonantly coupled to the cavity 
mode. Depending on the structure and the doping level of the quantum well, the 
transition can be of either {\em inter-band}~\cite{YuCardona} or {\em 
inter-subband} nature~\cite{LiuCapassoBook}: in the former case, it involves 
some electrons being promoted from the valence to the conduction band and is 
typically located in the near-infrared or visible domain. In the latter case, 
electrons (or holes) are promoted from a filled subband of the conduction (or 
valence) band to another, initially empty subband. For a strong enough 
light-matter coupling, photons can undergo several absorption-emission cycles 
before being lost. In this so-called {\em strong light-matter coupling} 
regime, the eigenmodes are superpositions of photonic and matter excitations: 
depending on the nature of the electronic transition involved, we usually 
speak of {\em interband polaritons} or {\em intersubband polaritons}.

As a few decades of intense theoretical and experimental research have shown, 
interband (IB) polaritons combine the extremely low mass of cavity photons 
(orders of magnitude lower than the excitonic one) with sizable interactions 
stemming from their excitonic component. These remarkable properties have been 
at the heart of the celebrated observations of Bose-Einstein 
condensation~\cite{Nature.443.409} and superfluidity 
effects~\cite{NaturePhys.5.805,Amo:2011Science} in a polariton gas and are 
still being widely exploited in a number of exciting directions, as reviewed 
in~\cite{Microcavities, RevModPhys.85.299, 1802.04173}.

Since with the early observations \cite{PhysRevLett.90.116401, 
PhysRevB.68.245320}, the research on intersubband (ISB) polaritons has also 
made impressive steps. In addition to their operation at much longer 
wavelengths in the mid or far infrared (MIR or FIR, respectively), ISB 
polaritons display a few other remarkable differences compared to IB ones. 
Since a large number of electrons are involved in the optical transition, the 
Rabi splitting can be pushed to extremely high values just by increasing the 
doping level in the well, in some cases even comparable to the transition 
frequency in the so-called {\em ultra-strong} light-matter coupling regime as 
predicted in~\cite{ciuti2005quantum,PhysRevB.85.045304} and experimentally 
observed 
in~\cite{PhysRevB.79.201303,PhysRevLett.105.196402,PhysRevLett.108.106402}. 
The strong coupling to the electromagnetic field displayed by ISB transitions 
is typically associated to much faster decay processes, which have so far 
limited the Q-factor of the polariton modes well below 100: while this is well 
enough to clearly observe the polariton mode 
splitting~\cite{PhysRevLett.90.116401, PhysRevB.68.245320, 
PhysRevLett.102.186402, PhysRevLett.105.196402}, relatively high pump 
intensities are needed to clearly observe nonlinear 
phenomena~\cite{PhysRevB.86.201302,PhysRevB.91.085308}. 
In spite of these difficulties, a wide community is actively involved in the 
study of nonlinear optical phenomena in these systems, from new sources of 
coherent light in the MIR/FIR domain based on parametric 
oscillation~\cite{PhysRevLett.102.136403,PhysRevB.87.235322}, Bose-Einstein 
condensation~\cite{PhysRevX.5.011031}, or inter-polariton 
transitions~\cite{PhysRevB.87.241304}, to the development of optical comb 
sources~\cite{Nature.492.229,OptExpress.23.5167}, to the generation of 
nonclassical states of light~\cite{ciuti2005quantum,Nature.458.178}.

Except for the above-cited works, the theoretical study of these processes is 
still lacking a microscopic model of interaction between ISB polaritons and an 
flexible and easily manageable theory to describe nonlinear processes with ISB 
polaritons in different configurations and geometries. While the former issue 
will be the subject of a forthcoming work~\cite{Leymann}, the goal of the 
present article is to introduce a generalized Gross-Pitaevskii equation (GGPE) 
to describe the nonlinear spatio-temporal dynamics of ISB polaritons. Even 
though our discussion will be focussed on a specific family of patterned 
cavities currently used in experiments~\cite{JMM2014,PhysRevB.96.235301}, our 
GGPE approach has a much wider application range to generic laterally 
patterned 
planar devices. 

Our theory is of course inspired by an analogous description of IB polaritons, 
but new technical difficulties originate in ISB systems from the much wider 
range of frequencies covered by the polaritons and the consequent importance 
of 
non-Markovian effects in the radiative decay. In contrast to IB polaritons 
whose radiative coupling is typically weakly dependent on frequency and 
momentum in the region of interest~\cite{RevModPhys.85.299}, special care is 
required in the ISB case to correctly enforce the light cone condition on the 
external propagating modes. This challenge is the main theoretical contribution 
of this work.

The power of our theory is confirmed by successfully comparing its predictions 
with experimental reflection spectra obtained in the linear 
regime~\cite{JMM2014,PhysRevB.96.235301}. As a first concrete application of 
the theory, we will take inspiration from related experiments with IB 
polaritons~\cite{PhysRevLett.84.1547, PhysRevB.62.R16247, 
SemicondSciTechnol.22.R1, RevModPhys.85.299} and study the possibility of 
optical parametric oscillation 
processes for ISB polaritons. The potential of these processes to realize 
novel sources of coherent light in the MIR and FIR domains is finally sketched.

% A consistent microscopic treatment of driven-dissipative effects is 
% therefore needed before the GGPE method can be applied to practical problems 
% of experimental relevance. 

The structure of the article is the following. In Sec.~\ref{sec:model} we 
review the physical system under consideration and we introduce the 
theoretical model based on the GGPE. The crucial novelty of the approach is 
summarized in Sec.\ref{sec:radiative}, which deals with the non-Markovian 
features appearing in the radiative coupling of ISB polaritons with the 
external radiation. The quantitative accuracy of the method is illustrated in 
Sec.\ref{sec:linear}, where we present computed reflectivity spectra in 
striking agreement with experimental observations presented in the recent 
literature. First applications of the GGPE approach to nonlinear problems are 
presented in the following Section: results for triply-resonant optical 
parametric oscillation are presented in Sec.\ref{sec:OPO}, while pump-probe 
optical amplification is studied in Sec.\ref{sec:pump_probe}. Conclusions and 
perspectives are finally summarized in Sec.\ref{sec:conclusions}.

\section{The physical system and its modeling}
\label{sec:model}

\begin{figure}[tb!]
 \includegraphics[width=5cm]{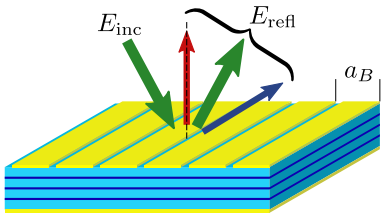}
\caption{(Color online) Sketch of the microcavity device under consideration. 
A series of quantum wells (dark blue) is embedded in a planar microcavity 
device enclosed by a pair of metallic mirrors (yellow). The top mirror is 
laterally patterned into a grating to ensure radiative coupling of the cavity 
modes with the external radiation. The device is operated by shining a pump 
beam of electric field $E_\mathrm{inc}$ at an angle to its surface. The 
reflected field $E_\mathrm{refl}$ is dominated by the specular reflection of 
the pump, but nonlinear phenomena can give rise to other spectral components.}
\label{fig:device_diagram}
\end{figure}

The physical system under consideration is sketched in 
Fig.~\ref{fig:device_diagram}. It consists of a solid state device featuring a 
series of two dimensional quantum wells (QW) embedded in a planar 
semiconductor microcavity. The cavity mode is enclosed by a pair of 
plane-parallel metallic mirrors and is able to propagate along the $xy$ cavity 
plane. Since we are interested in maximizing the interaction to the ISB 
transition, we focus our attention on TM-polarized electromagnetic modes with a 
uniform profile along $z$ and a wavevector parallel to the cavity plane. This 
allows to have an electric field fully polarized along the $z$ direction of the 
intersubband dipole. In contrast to the case of distributed Bragg reflector 
microcavities used for IB polaritons, the cavity modes under examination here 
have a linear, massless dispersion
\begin{equation}\label{eq:dispersion_photon}
\omega_C(\kk) = c\,|\kk|/n
\end{equation}
with a background refractive index $n$ of the cavity material on the order of 
$n\approx 3$ for typical samples. This means that the photon dispersion in the 
unpatterned cavity lies below the light cone $\omega=c |\kk|$ and is thus 
decoupled from external radiative modes. 

In order to overcome this limitation, the front mirror is laterally patterned 
as a Bragg grating along $x$. The spatial periodic modulation along the cavity 
plane induces a Bragg-folding of the dispersion, which enables an effective 
coupling to the external radiative modes outside the cavity. The period of the 
grating selects the operation wavelength: in the 
experiments~\cite{PhysRevLett.102.186402,JMM2014,PhysRevB.96.235301}, the 
operation wavelength is located in the neighborhood of the second Bragg gap 
which opens around $\kk=0$. A typical experiment then consists of sending one 
(or more) incident beam(s) onto the device and collecting the the light that is 
elastically reflected by the system and/or the new frequency components that 
are generated by the optical processes taking place inside the cavity. As usual 
in planar microcavities, the in-plane wavevector of the polaritons directly 
reflects into the angle made by the external radiation to the 
normal~\cite{RevModPhys.85.299}.

Within the wavevector range of optical interest, ISB transitions are 
nondispersive, at a momentum-independent frequency $\omega_X= \wisb$. Since 
they involve the motion of charged carriers, they are affected by Coulomb 
interactions between carriers. This is expected to result in strong two-body 
interactions between intersubband excitations, which then translate into 
efficient two-body scattering processes between polaritons or, in optical 
terms, sizable $\chi^{(3)}$ nonlinear optical susceptibilities. 

\subsection{Conservative light-matter coupling: intersubband polaritons}

The dynamics of the photonic cavity field $\psic(\rr)$ and of the ISB excitonic 
field $\psix(\rr)$, and the strong light-matter interactions occurring between 
the two, can be cast in terms of a pair of coherently coupled generalized 
Gross-Pitaevskii equations (GGPE)~\cite{RevModPhys.85.299}, extending to 
the nonequilibrium polariton case the formalism used in the context of 
Bose-Einstein condensates of ultracold 
atoms,
\begin{widetext}
\begin{align}
 i\cdiff{t} \psic(\rr) &=
     \left[ \omega_C(-i\nabla) -\frac{i}{2}\gamma_{\mathrm nr} \right] 
            \psic(\rr)
    + V(x) \psic(\rr)
    + \Omega_R \psix(\rr)
    - \Gamma_\mathrm{rad}(\rr,t) + \Pi(\rr, t), \label{eq:gpe_c}\\
i\cdiff{t} \psix(\rr) &=
    \left[ \omega_X - \frac{i}{2} \gamma_h \right] \psix(\rr)
    + \frac{g}{2}|\psix(x)|^2 \psix(\rr)
    + \Omega_R \psic(\rr)\,. \label{eq:gpe_x}
\end{align}
\end{widetext}
The nonradiative losses of the cavity mode and the homogeneous losses of the 
ISB excitations are represented by the coefficients $\gamma_\mathrm{nr}$ and 
$\gamma_\mathrm{h}$, respectively, and they are assumed to be local and 
homogeneous in space. The strength of the coherent light-matter coupling is 
quantified by the Rabi frequency $\Omega_R$. To reduce the technicalities, we 
focus our attention here on the case of a strong but not ultra-strong 
coupling, in which $\omega_{X,C} \gg \Omega_R \gg \gamma$.

Since a complete theory of interactions between intersubband excitations is 
still in the course of being developed~\cite{Leymann}, these are modeled here 
in the simplest way as spatially local binary interactions of strength $g$. 
Since the value of this quantity is still unknown, throughout this work we 
will express interaction energies in terms of the experimentally observable 
blue-shift $g\,|\psi_X|^2$.

The effect of the Bragg grating is the most subtle one and is responsible for 
the main differences compared to the case of IB polaritons. On one hand, it 
introduces into the GGPE an external potential $V(x)$ term acting on the 
photonic field, diagonal in coordinate space and periodic with period $a_B$. 
On the other hand, new Bragg-scattering-mediated radiative channels exists, 
through which light can be injected (extracted) into (from) the cavity.

\subsection{Effect of pump and losses}
\label{sec:radiative}

The last two terms in Eq.~\ref{eq:gpe_c} account for the radiative losses 
($\Gamma_\mathrm{rad}$) and the external pumping of the cavity field ($\Pi$). 
In the case of IB polaritons all $\kk$-modes involved in the dynamics are 
radiative and can be described~\cite{RevModPhys.85.299} within a Markovian 
theory in terms of a (almost) frequency- and momentum-independent radiative 
loss rate $\gamma_C$. Here, the Bragg-folded cavity dispersion crosses the 
light cone, which requires the frequency-dependence to be explicitly included 
and gives rise to non-Markovian features in the decay process. An efficient 
modeling of this physics in the GGPE framework is the subject of this section.

\begin{figure}[tb!]
\centering
\includegraphics{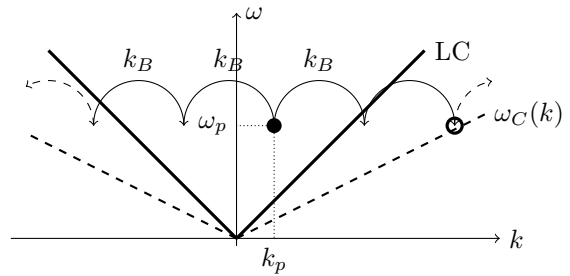}
\caption{Sketch of the Bragg scattering-mediated coupling of cavity modes to 
external radiation. An incident photon (full circle) of in-plane momentum 
$k_p$ and frequency $\omega_p$ must lie within the radiative region above the 
lightcone (LC) indicated by the full line. The incident light wavevector 
changes by multiples of the Bragg wave number $k_B$ as a consequence of the 
scattering on the grating, thus allowing to resonantly couple to the cavity 
mode (open circle).}
\label{fig:bragg_scattering}
\end{figure}

The incident field $E_\mathrm{inc}(\kk, \omega)$ only involves propagating 
radiative modes satisfying the light-cone condition $\omega \geq c\,|\kk|$. 
Even though no direct resonant coupling to the optical modes of an unpatterned 
planar cavity can occur, Bragg scattering processes on the grating are 
effective in relaxing this condition, as sketched in 
Fig.\ref{fig:bragg_scattering}. This can be formally captured by the relation 
\begin{align} \label{eq:pump_convolution}
    \tilde{\Pi}(\kk, \omega) = \int \mathcal{T}(\kk, \kk') 
                               \tilde{E}_\mathrm{inc}(\kk', \omega) dk',
\end{align}
between the Fourier components of the in-cavity and radiative fields, where the 
convolution kernel $\mathcal{T}(\kk, \kk')$ acts as a generalised transmission 
coefficient. The conjugate of the same transmission coefficient appears in the 
expression 
\begin{equation}
\tilde{E}_\mathrm{refl}(\kk', \omega) =
                \tilde{E}_\mathrm{inc}(\kk', \omega) 
                    -i \mathcal{T}^*(\kk, \kk') \tilde{\psic}(\kk, \omega)\,.
\end{equation}
for the radiation emerging from the cavity -- also this latter being of course 
localized in the radiative region $\omega \geq c\,|\kk|$.

\begin{figure}[bt!]
\includegraphics[width=8.5cm]{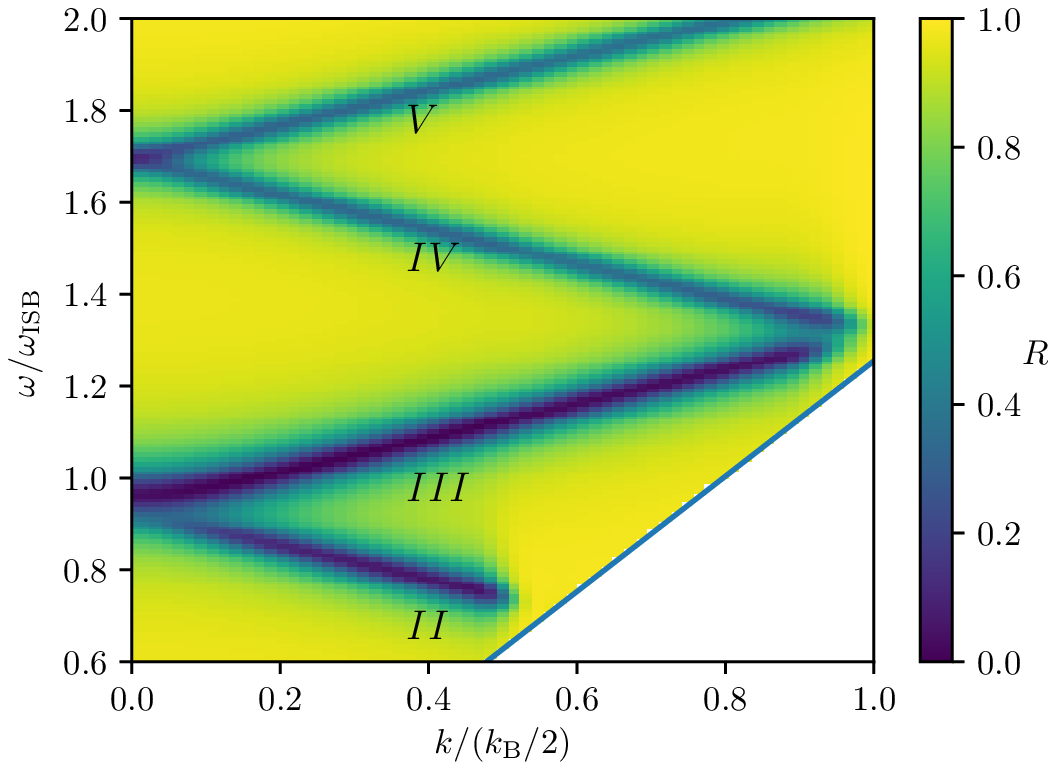}
\includegraphics[width=8.5cm]{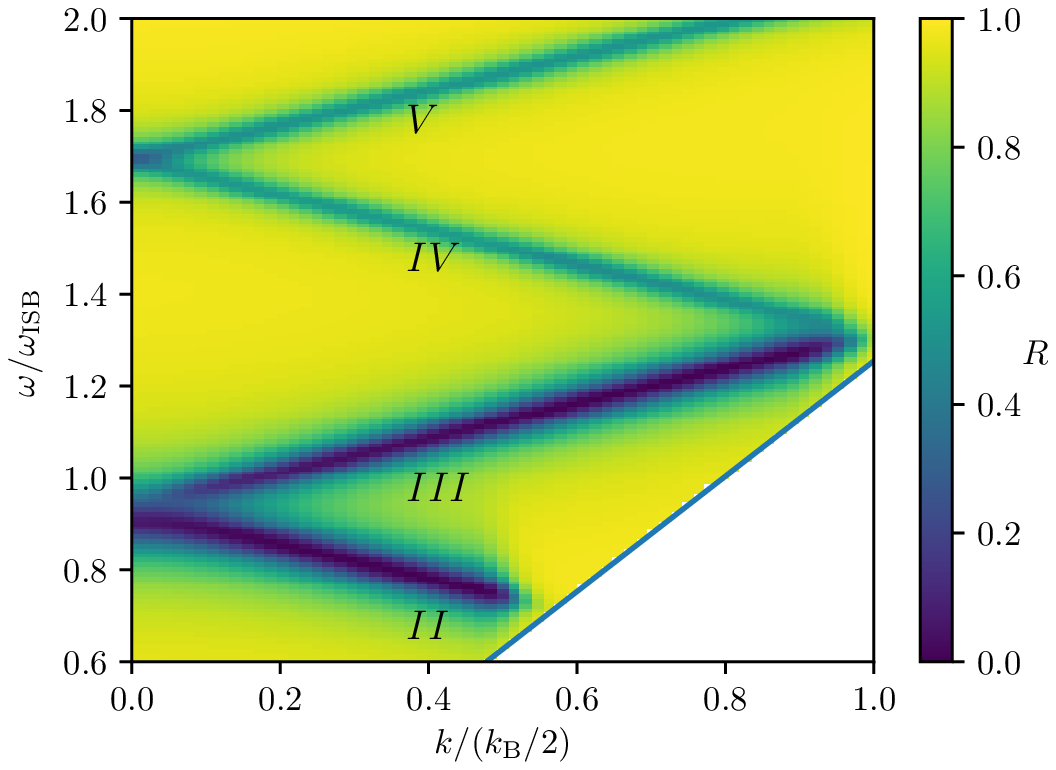}
 \caption{(Color online) Colorplots of the frequency- and momentum-dependent 
reflectivity $R(\kk,\omega)$ for emtpy cavities, i.e., $\Omega_R=0$. The Bragg 
grating consists of a square wave potential of period $a = 
\unit{4.26}{\micron}$ with filling factor 75\% (top panel, comparable to the 
device used in Ref.~\cite{PhysRevB.96.235301}) and 25\% (bottom panel). The 
remaining simulation parameters are reported in Appendix~\ref{app:parameters}. 
Roman numerals label the order of each band. The different visibility of 
the bands is the sole consequence of the different filling factor. The 
diagonal solid line indicates the light-cone condition: the white region below 
it in the bottom-right direction is inaccessible to reflection experiments.}
\label{fig:bare_cavity}
\end{figure}

While the form of these equations is fixed by the general structure of the 
problem, the detailed form of $\mathcal{T}(\kk, \kk')$ depends on the 
microscopic details of the sample. Given the discrete translational symmetry 
of the grating, $\mathcal{T}(\kk, \kk')$ will consist of a series of 
$\delta$-peaks at $k_y=0$ and $k_x= m k_B$ (with integer $m$), i.e. integer 
multiples of the Bragg wavevector $k_B=2\pi/a_B$. One can also anticipate that 
the amplitude of the different $\delta$-peaks is a decreasing funcion of 
$|\kk-\kk'|$ and roughly proportional to the Fourier transform of the 
potential $V(x)$ experienced by the cavity mode in Eq.~\ref{eq:gpe_c}, i.e., 
\begin{equation}
\mathcal{T}(\kk, \kk') \propto \tilde{V}(|\kk - \kk'|),
\end{equation}
with the proportionality factor set phenomenologically to reproduce the 
physical linewidth of a reference cavity band. In our calculations, we do 
this with respect to the third photonic band (cf.~Fig.~\ref{fig:bare_cavity}). 

More care is needed for the radiative loss term $\Gamma_{\rm rad}$. This is 
once again due to the strong frequency dependence of the spectral density of 
the radiative modes, that are spectrally localized above the light-cone $\omega 
\geq c\,|\kk|$. Below it, the spectral density of radiative modes is rigorously 
zero. This introduces non-Markovian features in the dissipative 
dynamics~\cite{Petruccione}.

If both the incident and the cavity fields are monochromatic at frequency 
$\omega$, it is enough to evaluate the decay at the given frequency $\omega$, 
so that radiative losses can be directly included into the GGPE as temporally 
local $\kk$-space terms 
\begin{align}
    \tilde{\Gamma}_\mathrm{rad}(\kk,t) = - \frac{i}{2} 
            \int\! d\kk''\, \Gamma_\omega(\kk, \kk'')\,\psic(\kk'',t)\,d\kk'',
\end{align}
where the integral in the kernel
\begin{equation} \label{eq:input-output}
\Gamma_\omega(\kk, \kk'') = \int_{|\kk'|\leq \omega/c} d\kk'\, 
                        \mathcal{T}^*(\kk'', \kk')\,\mathcal{T}(\kk, \kk')
\end{equation}
runs over those intermediate radiative modes $\kk'$ lying above the 
light-cone. The small shift of the cavity modes due to the coupling to the 
radiative modes outside the cavity has been neglected. In its simplicity, this 
exactly solvable single-frequency case encompasses polariton superfluidity 
experiments~\cite{NaturePhys.5.805} and quantum hydrodynamic soliton 
nucleation~\cite{Amo:2011Science}: whereas this physics has been so far 
studied only in the exciton-polariton case, our theory provides an efficient 
tool to investigate it in the new case of ISB polaritons.

For a general dynamics, the frequency-dependence of the radiative decay leads 
to a temporally non-local response,
\begin{align}
    \tilde{\Gamma}_\mathrm{rad}(\kk,t) = - \frac{i}{2} \int d\kk'' 
\int_0^\infty\!d\tau\,\bar{\Gamma}(\kk, \kk'';\tau)\,\psic(\kk'',t-\tau),
\end{align}
where the delay kernel $\bar{\Gamma}(\kk, \kk'';\tau)$ is the inverse Fourier 
transform of $\Gamma_\omega(\kk, \kk'')$ with respect to $\omega$, and 
accounts for both dissipative and reactive effects of the coupling to the 
radiative modes \cite{cohen1998}. Since the 
temporal non-locality of $\bar{\Gamma}$ introduces technical difficulties in 
the solution of the GGPE Eqs.~\ref{eq:gpe_c}-\ref{eq:gpe_x}, for practical 
problems it is useful to devise approximate treatments that provide accurate 
results while preserving temporal locality.

Unless one is dealing with features lying near to the edge of the light cone, a 
natural strategy to this purpose is the approximate $\Gamma_\omega(\kk, \kk'')$ 
with its value at the bare mode frequencies. In our case, it is convenient to 
choose $\omega=c(|\kk|+|\kk'|)/2$, which has the advantage of being symmetric 
under exchange of $\kk$ and $\kk'$ and to recover the common value when the 
frequencies of the two modes coincide and the damping term is most effective. 
In the next sections we will make use of this prescription and we will see that 
it provides accurate and predictive results.

Our numerical implementation uses the so-called split-step method to include 
the potential and the kinetic energy in the GGPE as diagonal terms in either 
real- or momentum-space, and to use a standard Fast Fourier transform method to 
quickly switch between them. The pump and radiative damping terms, which are 
not diagonal in neither real- nor momentum-space, are implemented by first 
order discretisation of time at the end of  each split-step evolution interval. 
To reduce the numerical load, the simulations are carried out in the rotating 
frame in which the carrier pump mode is stationary. 

%We assume that each space interval $\Delta x$ corresponds to a 
%two-dimensional tile of area $\Delta x \cdot \lambda_\mathrm{ISB}$, with  
%$\lambda_\mathrm{ISB}$ the length scale of the simulation. In this way, the 
%length of the one dimensional simulation domain and the physical two 
%dimensional surface area share the same numerical value. The pump field is  
%normalised such that, over a time-step $\Delta t$, the energy deposited on  
%each tile over an evolution time step $\Delta t$ is equal to $I \Delta \Delta 
%x^2$, with $I \propto E^2$ the physical pump intensity.

The convergence of the results was assessed by simulating the same system on 
progressively finer meshes. As a rule of thumb, it was found that results are 
converged when $a_B / \Delta x \gtrsim 8$.

In the definition of the matrix $\mathcal{T}(\kk, \kk')$ there are two 
implementation subtleties that the reader should be aware of. First, the 
spectrum of the Bragg potential, $\tilde V(\kk)$, must decay sufficiently fast 
to zero for large 
wavenumbers in order for the simulation to converge as the mesh is refined. 
Depending on  the particular Bragg pattern, it may be convenient to smoothen 
down large wavenumber components of the potential according to an exponential 
law. Second, the lightcone condition introduces sharp edges, which may give 
rise to high frequency and high wavenumber artifacts. Once again, the solution 
is to smoothen the edge of the lightcone according to an exponential law.

In order to simplify the technicalities, we assume that the grating is 
homogeneous along the $y$ direction and we focus our attention on light fields 
at $k_y=0$. Under these assumptions, a one-dimensional version of the GGPEs 
can be used. If needed, no conceptual difficulty would prevent our results 
from being straightforwardly extended to two dimensional patterns and 
arbitrary in-plane wavevectors and incidence angles.

\section{Linear Optics}
\label{sec:linear}

For sufficiently weak driving, we can assume that linear response is a 
satisfactory description. Indeed, without a sufficient build up of the ISB 
excitonic modes, the contribution $g|\psix|^2$ to the total energy is 
negligible when compared to the other energy scales at play in 
Eq.~\ref{eq:gpe_x}. 

Reflection measurements are a powerful tool to gain knowledge of the physics 
taking place inside the microcavity. The reflected intensity 
$I_\mathrm{refl}(\kk, \omega) = |\tilde{E}_\mathrm{refl}(\kk, \omega)|^2$ is a 
relatively easy measurement in experiments. The in-plane wavenumber can be 
selected by changing the angle at which one collects (shines) the reflected 
(incident) radiation, and channeling the radiation through a Fourier transform 
spectrometer gives access to the spectral distribution.

Referencing the reflected field to the incident one, we recover the usual 
definition of the reflectivity,
\begin{equation}
 R(\kk, \omega) = \frac{I_\mathrm{refl}(\kk, \omega)}{I_\mathrm{inc}(\kk, \omega)}
              = \frac{|\tilde{E}_\mathrm{refl}(\kk, \omega)|^2}
                     {|\tilde{E}_\mathrm{inc}(\kk, \omega)|^2},
\end{equation}
which we are going to use extensively in the following sections. Plots of the 
reflectivity in different cases are shown in Fig.~\ref{fig:reflectivity}. 

A more in depth experimental investigation can benefit from electro-optic 
sampling techniques, allowing the real time measurement of the actual 
reflected electric field $\tilde{E}_\mathrm{refl}(\kk, t)$ 
\cite{ApplPhysLett.71.1285, ApplPhysLett.76.3191, ApplPhysLett.85.863, 
ApplPhysLett.85.3360, OptLett.8.2435, NaturePhotonics.10.159, 
OptLett.21.4367}. The direct measurement of both amplitude and phase of the 
reflected radiation allows to study more subtle optical features such as the 
temporal shape of wavepackets or the coherence properties of the emission.

\subsection{Bare cavity}

It is instructive to first consider the case of an empty cavity, in which no 
ISB transition is present or, equivalently, this is decoupled from the cavity 
field $\Omega_R=0$. Because of the grating, the linear dispersion of the free 
photon is folded around the first Brillouin zone and gaps open because of 
Bragg scattering processes. 

Examples of reflectivity spectra in the $(k_x,\omega)$ plane at $k_y=0$ for an 
empty cavity and two different gratings are shown in  
Fig.~\ref{fig:bare_cavity}. In the reflectivity plots, the photonic bands show 
up as reflectivity minima, or, in other words, as peaks in the resonant 
absorption of the incident field.

The physical parameters of the calculation were chosen so as to closely match 
the devices used in current experiments~\footnote{To obtain a better match with 
experimental data, we had to slightly modify the photon dispersion relation in 
the region of interest to 
\begin{equation}
\omega_C(k) = |k|/n + \omega_0,
\end{equation}
with a small $\omega_0 = \unit{20}{\meV}$, so to include the consequences of, 
e.g., the leakage of electromagnetic radiation inside the metallic mirrors.}, 
cf.~App.~\ref{app:parameters}.
The grating is included in the GGPE as
\begin{equation}
 V(x) = V_0 \begin{cases}
             2-2\nu &\mathrm{if}\ 0   \leq x\ \mathrm{mod\ } a_B \leq 
\nu a_B, \\
             -2\nu  & \mathrm{if}\ \nu a_B < x\ \mathrm{mod\ } a_B \leq 1,
            \end{cases}
\end{equation}
i.e., a zero-mean square wave of amplitude $V_0$, period $a_B$ and filling 
factor $\nu$. The results in Fig.~\ref{fig:bare_cavity} only differ by the 
filling factor of the grating, namely, $\nu = 75\%$ and $\nu = 25\%$ for top 
and bottom panels, respectively, the former being representative of the 
experimental conditions~\cite{JMM2014,PhysRevB.96.235301}.
The visibility of the different bands changes with the details of the Bragg 
grating. For instance, the second band tends to fade towards $k=0$ and the 
third band is more pronounced at high filling factors, whereas the opposite is 
true for low filling. In the symmetric case (50\% filling, not shown), both 
bands are approximately equally visible.

This shows that our modelling of the pumping and of the radiative losses in 
terms of the transmission matrix $\mathcal{T}(\kk, \kk')$ as described in 
Sec.\ref{sec:radiative} is able to capture the nuances and complexities of the 
grating potential without the need to manually tune any parameter, as it was 
instead the case in the temporal coupled-mode theory of 
Ref.~\cite{PhysRevB.96.235301}. As a result, more complicated gratings can be 
studied without the need to modify the general structure of the model, which 
can be readily extended to fully two dimensional geometries, too.

\subsection{Strong coupling regime}

\begin{figure}[bt!]
 \includegraphics[width=8.5cm]{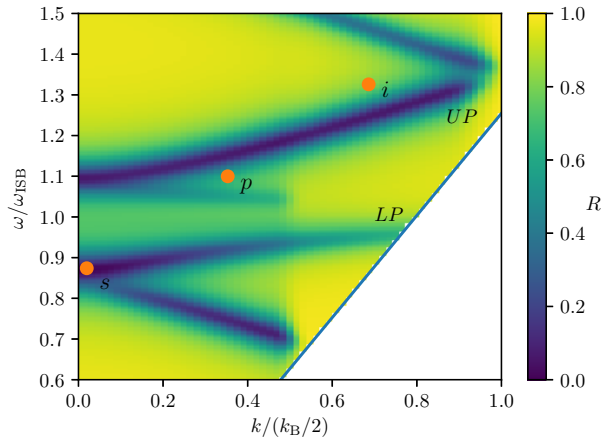}
\caption{(Color online) Reflectivity $R(\kk, \omega)$ of cavity modes strongly 
coupled to an intersubband transition. The simulation parameters are reported 
in Appendix~\ref{app:parameters}. The dots indicate the pump, signal, and 
idler modes of the OPO discussed in Sec.\ref{sec:OPO}.}
% {\bf value of $\Gamma_{\rm rad}$... Color jump at $k\approx 
%0.5$ on right panel...} }
 \label{fig:reflectivity}
\end{figure}

Still in the weak pump limit, we now allow for nonzero values of the Rabi 
coupling $\Omega_R$, so that the system can enter the strong coupling regime. 
The reflectivity for this case is displayed in Fig.\ref{fig:reflectivity}. As 
expected, there is a clear avoided crossing between the non-dispersive ISB 
transition at $\wisb$ and the bare cavity photonic dispersion, with the 
appearance of the typical upper and lower polariton branches separated by a 
gap of the order of $2\Omega_R$. In between the two main polariton bands 
indicated as LP and UP in Fig.~\ref{fig:reflectivity}, weak additional bands 
of dominant ISB character appear in the energy region around $\wisb$. Their 
physical origin is due to the Bragg folding of the main polariton branches.

The accuracy of this approach is apparent when this plot for the reflectivity 
is compared to experimental measurements reported, e.g., in 
Ref.~\cite{JMM2014,PhysRevB.96.235301}. In contrast with the 
phenomenological temporal coupled-mode theory used in the works cited, we 
provide a more microscopic treatment, which can be straightforwardly extended 
to more complex geometries as well as to the far richer nonlinear physics in 
the presence of interactions between polaritons. This topic is the subject 
of the next Section.

\section{Nonlinear optics}
\label{sec:nonlinear}

\begin{figure}
\begin{center}
\includegraphics[width=8.5cm]{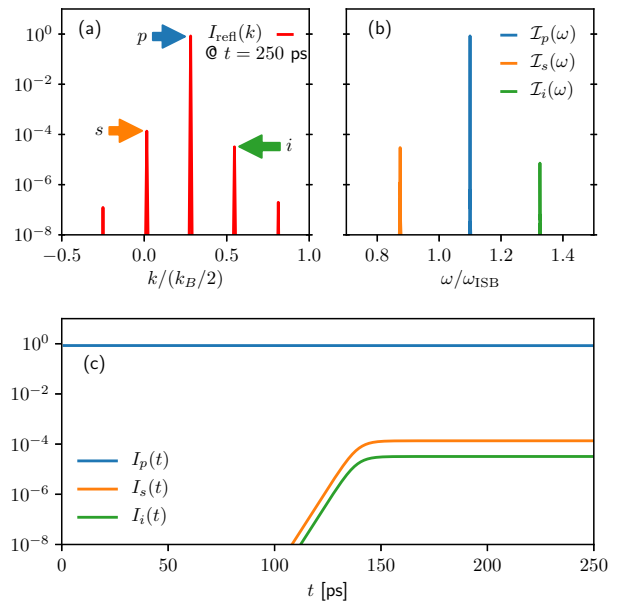}
%img_realt_D0.54528_N1024_n3.3_woff20.0_ISB116.0_W13.0_gcn5.40_gcr4.50_gx6.0
%0_s3e-05_V10.0_a4.26_f0.25_F0.500_w1.1000_k0.352827_pnoise1e-10
\end{center}
\caption{(Colour online) Optical parametric oscillation under continuous-wave 
pumping at $k_p / (k_B/2) \approx 0.28$, $\omega_p / \wisb = 1.1$ and the 
simulation parameters of App.~\ref{app:parameters}. The positions of the pump, 
signal and idler are marked by dots in Fig.~\ref{fig:reflectivity}. The pump 
intensity and nonlinear coupling are tuned such that $g|\psix|^2 = 
\unit{23.1}{\meV}$. All intensities are normalised to the intensity of the 
pump mode. (a) Steady state (in practice computed after \unit{250}{\ps} of 
evolution) $k$-space momentum distribution of the emitted radiation. Besides 
the signal and idler peaks, one sees further replicas due to higher order 
diffraction from the grating. (b) Momentum-selected (i.e., angle-selected) 
emitted radiation spectrum in the frequency domain, $\mathcal{I}_{\{p, s, 
i\}}(\omega) = I(k_{\{p, s, i \}}, \omega)$ (c) Intensities of pump, signal 
and idler emission in real time, $I_{\{p, s, i\}}(t) \equiv 
I_\mathrm{refl}(k_{\{p, s,i\}}, t)$}
\label{fig:opo_mir}
\end{figure}

As the intensity of the incident beam is increased, nonlinear phenomena start 
to play a crucial role in the dynamics. Nonlinearities, arising from 
polariton-polariton scattering---or from other channels not discussed in the 
present paper---will in general shift the energy position of the bands in a 
nontrivial way. The resonance condition between the incident pump and the 
energy shifted bands, and the in-cavity mode mixing and frequency conversion 
processes act together in a complex way, giving origin to a variety of 
nonlinear effects. A review of the most significant nonlinear effects in the 
context of IB exciton polaritons is given in Ref.\cite{RevModPhys.85.299}.

\subsection{Optical parametric oscillation}
\label{sec:OPO}

If the pump parameters are carefully tuned to resonantly excite suitable 
points $(\kk_p, \omega_p)$ on the polaritonic bands, the system may enter the 
so-called optical parametric oscillation (OPO) regime. In this regime, two pump 
polaritons are resonantly scattered to another pair of states, commonly known 
as \emph{signal} and \emph{idler} (hereafter noted with subscripts $s$ and 
$i$, respectively). The peculiar shape of the polaritonic band structure may 
in fact allow for the scattering process to be triply-resonant, with pump, 
signal and idler all lying on the dynamically shifted polaritonic bands. 
In this case, energy and momentum are conserved during the elementary 
parametric scattering process underlying OPO,
\begin{equation}
 \kk_i + \kk_s = 2 \kk_p, \qquad
 \omega_i + \omega_s = 2\omega_p.
 \end{equation}
As typical of bosons, a nonvanishing population in the final states further 
stimulates the scattering process. The signal and idler states may then become 
macroscopically occupied if the pump is strong enough to overcome losses, and 
may emit coherent radiation, normally with a narrow distribution in both 
wavenumber (i.e., angle) and frequency~\cite{RevModPhys.85.299}. The main 
result of this section is that the ISB polariton system can support polariton 
OPO operation.

While excitonic polaritons are studied in the visible or near-IR ranges, and 
are bound to the vicinity of the electronic band gap of the material, ISB 
systems allow for OPO operation in the novel spectral windows of MIR and THz 
frequencies, with a very wide tunability range: the Bragg frequency $ck_B$ of 
the cavity is simply determined by the grating period, while $\wisb$ is 
controlled by the QW thickness. This flexibility will be of paramount 
technological importance when one is to design OPO sources operating at a 
given desired frequency. To facilitate the application of our results to 
different frequency ranges, our results will be plotted in units of $\wisb$.

Figure \ref{fig:opo_mir}(a, b) show the emission spectra as predicted via a 
numerical
simulation of the coupled GGPEs (\ref{eq:gpe_c}-\ref{eq:gpe_x}) using realistic 
physical parameters. The drive is a monochromatic 
pump at energy $\omega/\wisb = 1.1$ and $23.3^\circ$ angle of incidence, i.e., 
wavenumber $k_p \approx 0.28 (k_B/2)$. An extremely weak random noise was 
added to the simulation to seed the scattering process. The signal and idler 
are emitted at $k_s \approx 0.02$ and $k_i \approx 0.55$ in units of $k_B/2$, 
at frequencies $\omega_s \approx 0.87\, \wisb$ and $\omega_i \approx 1.33\, 
\wisb$, respectively.
The triplet of pump, signal and idler obtained from the simulation is also 
marked with dots in the linear regime band diagram of 
Fig.~\ref{fig:reflectivity}. 
Their slightly shifted position out of resonance from the band is due to the 
significant blue shift of the upper polaritonic band due to the same 
polariton-polariton interactions that are responsible for the parametric 
scattering process.

As shown in Fig.~\ref{fig:opo_mir}(c), the emission from the signal and idler 
states grows in intensity over approximately \unit{150}{ps}, after which the 
system reaches a steady OPO state. However, this time scale wildly varies with 
the fine details of the system. Due to the stimulated character of the 
scattering process, the time required to reach the steady state is sensitive 
to the amount of initial noise that seeds the scattering process, and the 
actual nature and strength of the nonlinear coupling. Because of these 
limitations, we cannot assess the OPO threshold in terms of a physical pump 
intensity. Noneless, this can be estimated in terms of the nonlinear frequency 
shift $g\,|\psix|^2$, and is of the order of a few decay rates. For the 
realistic parameters of~\cite{PhysRevB.96.235301}, this would amount to around 
$\unit{15}{\meV}$. A specific study of the polariton interaction constant $g$ 
for different QW shapes and different levels of electronic doping is subject of 
on-going work~\cite{Leymann}: its main result will be a quantitative prediction 
for the value of the incident pump intensity at the OPO threshold.

\subsection{Optical amplification in pump-probe configuration} 
\label{sec:pump_probe}

\begin{figure}[bt!]
\includegraphics[width=6.5cm]{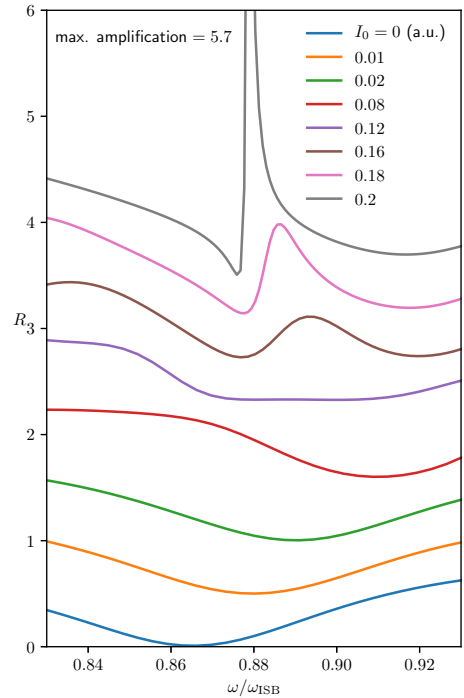}
\caption{Emission spectra with continuous wave pump ($\omega / \wisb = 
1.065$, $k / (k_B/2) \approx 0.28$) and \unit{150}{\fs} long Gaussian probe 
pulses ($\omega/\wisb = 0.89$, $k=0$) of growing intensity $I_0$, always below 
the OPO threshold located at $I_0^{\rm thr} \approx 0.22$ in these units. The 
probe intensity is taken very weak at $10^{-10}$. For clarity, each curve 
is offset by 0.5 from the previous one.}
\label{fig:pump_probe_below}
\end{figure}

Even below the threshold for a fully developed OPO, polariton-polariton 
scattering is able to transfer polaritons from the pump state to the signal 
and idler states. This phenomenon can be used to achieve optical amplification 
in a standard pump-probe configuration. An intense beam, ideally spectrally 
narrow, resonantly pumps the system, while a weaker probe beam, often 
spectrally broad, is shone so as to populate the final signal state, thus 
stimulating the scattering process. The onset of optical amplification is then 
witnessed by an intense narrow peak within the otherwise broad probe spectrum.

\subsubsection{Continuous wave pump}

We first consider the ideal case of a continuous wave pump below the threshold 
intensity of OPO. Figure~\ref{fig:pump_probe_below} shows the $k=0$ emission 
spectra for increasing pump intensity. A monochromatic pump ($\omega/\wisb = 
1.065$, $k/(k_B/2) \approx 0.28$) is paired with a \unit{150}{\fs} Gaussian 
probe pulse at $k=0$, centered at $\omega/\wisb = 0.89$. For this choice of 
parameters, the probe is approximately resonant with the lower polariton 
branch, yet with a much broader linewidth. Its intensity is extremely weak, to 
guarantee that the dynamics is deeply in the linear regime with respect to the 
probe amplitude.

As the pump intensity $I_0$ is ramped up, the first feature found in the 
numerics is a blue shift of the lower polariton minimum from its linear regime 
position (shown by the $I_0=0$ curve in the figure). Most remarkably, an 
experimental measurement of this lineshift as a function of the pump intensity 
would provide a useful estimate of the actual strength of the nonlinear 
coupling $g$ and of the OPO threshold intensity. Simultaneous recording of all 
polariton branches would shine light on the precise mechanism underlying 
polariton-polariton interactions~\footnote{Another pump-only strategy to 
estimate polariton interactions is of course based on the electro-optical 
sampling of the time-dependence of the emission frequency during ring-down 
oscillation after a strong pulsed pump.}
Further increasing the pump intensity, the emission spectra develop a peak 
that, sufficiently close to the threshold pump amplitude, becomes a sharp 
optical amplification signal.

We stress that an increase in the probe reflectivity at a given $(\kk,\omega)$ 
is not a firm evidence of a parametric scattering process. Such an increase 
can in fact originate from other effects, e.g., an increased polariton 
linewidth. Instead, the observation of a line narrowing effect 
and---especially---the direct detection of the third idler beam in addition to 
the pump and probe beams (typically at larger angles and frequencies) would be 
strong evidences that optical amplification is indeed taking place in the 
system.

\subsubsection{Pulsed pump}

\begin{figure}[tb!]
\begin{center}
 \includegraphics[width=8.5cm]{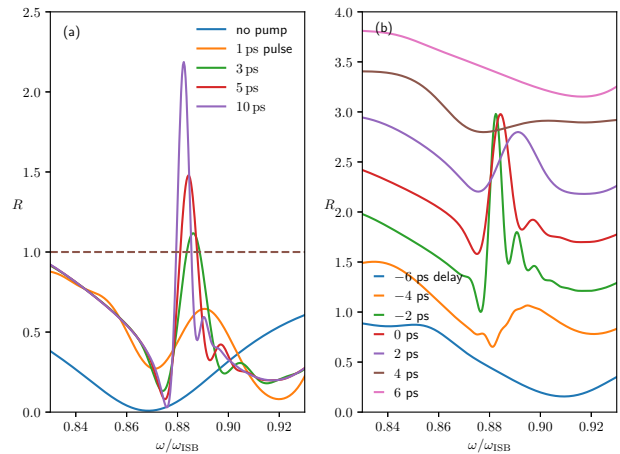}
\end{center}
\caption{Emission spectrum with pulsed pump of peak intensity $I_0 = 0.2$ 
in the units used in Fig.\ref{fig:opo_mir}, $\omega / \wisb = 1.065$, 
$2k / k_B \approx 0.28$) and \unit{150}{\fs} Gaussian pulse probe 
($\omega/\wisb = 0.89$, $k=0$, $I_0 = 10^{-5}$). (a) The pump is a Gaussian 
packet of increasing duration. The pump and probe pulses are synchronised so 
that they reach their maximum intensity at the same time. For pump pulses 
longer than approximately \unit{3}{\ps}, amplification is observed. (b) Effect 
of time delay between the probe and the pump pulse using under \unit{5}{\ps} 
pulsed pump. Curves are offset by 0.5 for clarity.}
\label{fig:pulsed_pump}
\end{figure}

Since no experimental measurement of the polariton-polariton interaction 
constant is available yet, we can not exclude that reaching the OPO threshold 
might require such high intensities that the only option could be the use of 
pulsed pump sources, all the way down to picosecond pulses or shorter. For such 
short pulses, peak intensities up to $\mathrm{GW}/\mathrm{cm}^2$ are in fact 
within reach of state-of-the-art sources.

In Fig.~\ref{fig:pulsed_pump}(a) we show the spectra obtained by driving the 
system with Gaussian pump pulses of increasing duration. The carrier frequency 
of the pump is the same of the previous subsection, as are the parameters of 
the probe. By stretching the pump duration, the system has time to build up 
population in the pump mode. When the pump is too short, parametric scattering 
processes do not have the time to efficiently amplify the probe. On the other 
hand, when the pump pulse is longer than approximately \unit{3}{\ps}, the 
scattering of polaritons from the pump to signal and idler states is strong 
enough to give an observable amplification of the probe.

This result should be seen as a proof of principle; given the very complex and 
sometimes even chaotic dynamics of OPOs, the time duration of the pump pulse 
required to reach actual amplification may depend on the fine details of the 
sample under consideration, as well as on the details of the pump-probe set-up 
and timing. As a general feature, we have observed a high sensitivity to the 
time delay between the arrival of the two pulses, as shown in 
Fig.~\ref{fig:pulsed_pump}(b). It appears that a slightly anticipation of the 
probe pulse with respect to the arrival of the pump, i.e., a negative delay, 
may result in higher amplification. This can be intuitively understood as a 
pre-seeding of the final signal state that has then time to develop during the 
whole time that the intensity of the pump is at its peak.

\section{Conclusions}
\label{sec:conclusions}
In this work we presented a new theoretical approach to study the linear and 
nonlinear dynamics of ISB polaritons resulting from the strong coupling 
between a planar cavity mode and an intersubband electronic transition in 
doped quantum wells. A special attention is paid to the non-Markovian 
driven-dissipative features occurring in the radiative coupling of the cavity 
polariton modes to the external radiation ones via Bragg scattering processes, 
which constitute the main difference from the well-known case of IB 
polaritons. The quantitative power of our approach is validated on the linear 
reflection spectra, that successfully reproduce the ones observed in recent 
experiments~\cite{JMM2014,PhysRevB.96.235301}. As a first example of concrete 
application, we studied optical parametric oscillation processes between 
intersubband polaritons and we outlined their perspectives in view of a new 
concept of coherent intersubband polariton laser sources 
\cite{PhysRevX.5.011031} operating in the MIR and FIR spectral domains. 

While experimental studies in this direction are actively in progress, new 
theoretical steps include a quantitative study of the microscopic interaction 
between polaritons~\cite{Leymann} and an extension of our theory to the 
ultrastrong coupling regime~\cite{ciuti2005quantum,Ciuti:2006PRA}, in which 
the non-rotating wave terms introduce substantial further complications in the 
treatment of the radiative decay channels.

We acknowledge financial support from the European Union FET-Open grant 
MIR-BOSE (737017) and from the Provincia Autonoma di Trento. We are grateful 
to C.~Ciuti, R.~Colombelli, M.~Knorr, C.~Lange, S.~De~Liberato and 
H.A.M.~Leymann for continuous stimulating discussions.

\appendix
\section{System parameters}
\label{app:parameters}
Unless otherwise stated, the system parameters used in the simulations are the 
following: $\wisb=\unit{116}{\meV}$, $n = 3.3$, $\omega_0 = \unit{20}{\meV}$, 
$\gamma_\mathrm{nr} = \unit{5.4}{\meV}$, $\gamma_\mathrm{r} = 
\unit{4.5}{\meV}$, $\gamma_\mathrm{h} = \unit{6.0}{\meV}$,  $V_0 = 
\unit{10}{\meV}$, $a_B=\unit{4.26}{\micron}$, filling factor $\nu = 75\%$ and 
$\Omega_R = \unit{13}{\meV}$. This choice of parameters is informed by the 
physical device used in the experiments of Ref.~\cite{PhysRevB.96.235301}.

\bibliography{biblio.bib}

\begin{thebibliography}{43}%
\makeatletter
\providecommand \@ifxundefined [1]{%
 \@ifx{#1\undefined}
}%
\providecommand \@ifnum [1]{%
 \ifnum #1\expandafter \@firstoftwo
 \else \expandafter \@secondoftwo
 \fi
}%
\providecommand \@ifx [1]{%
 \ifx #1\expandafter \@firstoftwo
 \else \expandafter \@secondoftwo
 \fi
}%
\providecommand \natexlab [1]{#1}%
\providecommand \enquote  [1]{``#1''}%
\providecommand \bibnamefont  [1]{#1}%
\providecommand \bibfnamefont [1]{#1}%
\providecommand \citenamefont [1]{#1}%
\providecommand \href@noop [0]{\@secondoftwo}%
\providecommand \href [0]{\begingroup \@sanitize@url \@href}%
\providecommand \@href[1]{\@@startlink{#1}\@@href}%
\providecommand \@@href[1]{\endgroup#1\@@endlink}%
\providecommand \@sanitize@url [0]{\catcode `\\12\catcode `\$12\catcode
  `\&12\catcode `\#12\catcode `\^12\catcode `\_12\catcode `\%12\relax}%
\providecommand \@@startlink[1]{}%
\providecommand \@@endlink[0]{}%
\providecommand \url  [0]{\begingroup\@sanitize@url \@url }%
\providecommand \@url [1]{\endgroup\@href {#1}{\urlprefix }}%
\providecommand \urlprefix  [0]{URL }%
\providecommand \Eprint [0]{\href }%
\providecommand \doibase [0]{http://dx.doi.org/}%
\providecommand \selectlanguage [0]{\@gobble}%
\providecommand \bibinfo  [0]{\@secondoftwo}%
\providecommand \bibfield  [0]{\@secondoftwo}%
\providecommand \translation [1]{[#1]}%
\providecommand \BibitemOpen [0]{}%
\providecommand \bibitemStop [0]{}%
\providecommand \bibitemNoStop [0]{.\EOS\space}%
\providecommand \EOS [0]{\spacefactor3000\relax}%
\providecommand \BibitemShut  [1]{\csname bibitem#1\endcsname}%
\let\auto@bib@innerbib\@empty
%</preamble>
\bibitem [{\citenamefont {Kavokin}\ \emph {et~al.}(2017)\citenamefont
  {Kavokin}, \citenamefont {Baumberg}, \citenamefont {Malpuech},\ and\
  \citenamefont {Laussy}}]{Microcavities}%
  \BibitemOpen
  \bibfield  {author} {\bibinfo {author} {\bibfnamefont {A.~V.}\ \bibnamefont
  {Kavokin}}, \bibinfo {author} {\bibfnamefont {J.~J.}\ \bibnamefont
  {Baumberg}}, \bibinfo {author} {\bibfnamefont {G.}~\bibnamefont {Malpuech}},
  \ and\ \bibinfo {author} {\bibfnamefont {F.~P.}\ \bibnamefont {Laussy}},\
  }\href {https://books.google.it/books?id=7EjADgAAQBAJ} {\emph {\bibinfo
  {title} {Microcavities}}}\ (\bibinfo  {publisher} {OUP Oxford},\ \bibinfo
  {year} {2017})\BibitemShut {NoStop}%
\bibitem [{\citenamefont {Yu}\ and\ \citenamefont {Cardona}(2005)}]{YuCardona}%
  \BibitemOpen
  \bibfield  {author} {\bibinfo {author} {\bibfnamefont {P.}~\bibnamefont
  {Yu}}\ and\ \bibinfo {author} {\bibfnamefont {M.}~\bibnamefont {Cardona}},\
  }\href {https://books.google.it/books?id=W9pdJZoAeyEC} {\emph {\bibinfo
  {title} {Fundamentals of Semiconductors: Physics and Materials Properties}}}\
  (\bibinfo  {publisher} {Springer Berlin Heidelberg},\ \bibinfo {year}
  {2005})\BibitemShut {NoStop}%
\bibitem [{\citenamefont {Liu}\ and\ \citenamefont
  {Capasso}(1999)}]{LiuCapassoBook}%
  \BibitemOpen
  \bibfield  {author} {\bibinfo {author} {\bibfnamefont {H.~C.}\ \bibnamefont
  {Liu}}\ and\ \bibinfo {author} {\bibfnamefont {F.}~\bibnamefont {Capasso}},\
  }\href {https://books.google.it/books?id=U6ceAQAAIAAJ} {\emph {\bibinfo
  {title} {Intersubband Transitions in Quantum Wells: Physics and Device
  Applications}}}\ (\bibinfo  {publisher} {Academic Press},\ \bibinfo {year}
  {1999})\BibitemShut {NoStop}%
\bibitem [{\citenamefont {Kasprzak}\ \emph {et~al.}(2006)\citenamefont
  {Kasprzak}, \citenamefont {Richard}, \citenamefont {Kundermann},
  \citenamefont {Baas}, \citenamefont {Jeambrun}, \citenamefont {Keeling},
  \citenamefont {Marchetti}, \citenamefont {Szymańska}, \citenamefont
  {André}, \citenamefont {Staehli}, \citenamefont {Savona}, \citenamefont
  {Littlewood}, \citenamefont {Deveaud},\ and\ \citenamefont
  {Dang}}]{Nature.443.409}%
  \BibitemOpen
  \bibfield  {author} {\bibinfo {author} {\bibfnamefont {J.}~\bibnamefont
  {Kasprzak}}, \bibinfo {author} {\bibfnamefont {M.}~\bibnamefont {Richard}},
  \bibinfo {author} {\bibfnamefont {S.}~\bibnamefont {Kundermann}}, \bibinfo
  {author} {\bibfnamefont {A.}~\bibnamefont {Baas}}, \bibinfo {author}
  {\bibfnamefont {P.}~\bibnamefont {Jeambrun}}, \bibinfo {author}
  {\bibfnamefont {J.~M.~J.}\ \bibnamefont {Keeling}}, \bibinfo {author}
  {\bibfnamefont {F.~M.}\ \bibnamefont {Marchetti}}, \bibinfo {author}
  {\bibfnamefont {M.~H.}\ \bibnamefont {Szymańska}}, \bibinfo {author}
  {\bibfnamefont {R.}~\bibnamefont {André}}, \bibinfo {author} {\bibfnamefont
  {J.~L.}\ \bibnamefont {Staehli}}, \bibinfo {author} {\bibfnamefont
  {V.}~\bibnamefont {Savona}}, \bibinfo {author} {\bibfnamefont {P.~B.}\
  \bibnamefont {Littlewood}}, \bibinfo {author} {\bibfnamefont
  {B.}~\bibnamefont {Deveaud}}, \ and\ \bibinfo {author} {\bibfnamefont
  {Le~Si}\ \bibnamefont {Dang}},\ }\bibfield  {title} {\enquote {\bibinfo
  {title} {{Bose–Einstein} condensation of exciton polaritons},}\ }\href
  {https://doi.org/10.1038/nature05131} {\bibfield  {journal} {\bibinfo
  {journal} {Nature}\ }\textbf {\bibinfo {volume} {443}},\ \bibinfo {pages}
  {409} (\bibinfo {year} {2006})}\BibitemShut {NoStop}%
\bibitem [{\citenamefont {Amo}\ \emph {et~al.}(2009)\citenamefont {Amo},
  \citenamefont {Lefr\`ere}, \citenamefont {Pigeon}, \citenamefont {Adrados},
  \citenamefont {Ciuti}, \citenamefont {Carusotto}, \citenamefont {Houdr\'e},
  \citenamefont {Giacobino},\ and\ \citenamefont {Bramati}}]{NaturePhys.5.805}%
  \BibitemOpen
  \bibfield  {author} {\bibinfo {author} {\bibfnamefont {A.}~\bibnamefont
  {Amo}}, \bibinfo {author} {\bibfnamefont {J.}~\bibnamefont {Lefr\`ere}},
  \bibinfo {author} {\bibfnamefont {S.}~\bibnamefont {Pigeon}}, \bibinfo
  {author} {\bibfnamefont {C.}~\bibnamefont {Adrados}}, \bibinfo {author}
  {\bibfnamefont {C.}~\bibnamefont {Ciuti}}, \bibinfo {author} {\bibfnamefont
  {I.}~\bibnamefont {Carusotto}}, \bibinfo {author} {\bibfnamefont
  {R.}~\bibnamefont {Houdr\'e}}, \bibinfo {author} {\bibfnamefont
  {E.}~\bibnamefont {Giacobino}}, \ and\ \bibinfo {author} {\bibfnamefont
  {A.}~\bibnamefont {Bramati}},\ }\bibfield  {title} {\enquote {\bibinfo
  {title} {Superfluidity of polaritons in semiconductor microcavities},}\
  }\href {https://doi.org/10.1038/nphoton.2015.269} {\bibfield  {journal}
  {\bibinfo  {journal} {Nature Physics}\ }\textbf {\bibinfo {volume} {5}},\
  \bibinfo {pages} {805} (\bibinfo {year} {2009})}\BibitemShut {NoStop}%
\bibitem [{\citenamefont {Amo}\ \emph {et~al.}(2011)\citenamefont {Amo},
  \citenamefont {Pigeon}, \citenamefont {Sanvitto}, \citenamefont {Sala},
  \citenamefont {Hivet}, \citenamefont {Carusotto}, \citenamefont {Pisanello},
  \citenamefont {Leménager}, \citenamefont {Houdré}, \citenamefont
  {Giacobino}, \citenamefont {Ciuti},\ and\ \citenamefont
  {Bramati}}]{Amo:2011Science}%
  \BibitemOpen
  \bibfield  {author} {\bibinfo {author} {\bibfnamefont {A.}~\bibnamefont
  {Amo}}, \bibinfo {author} {\bibfnamefont {S.}~\bibnamefont {Pigeon}},
  \bibinfo {author} {\bibfnamefont {D.}~\bibnamefont {Sanvitto}}, \bibinfo
  {author} {\bibfnamefont {V.~G.}\ \bibnamefont {Sala}}, \bibinfo {author}
  {\bibfnamefont {R.}~\bibnamefont {Hivet}}, \bibinfo {author} {\bibfnamefont
  {I.}~\bibnamefont {Carusotto}}, \bibinfo {author} {\bibfnamefont
  {F.}~\bibnamefont {Pisanello}}, \bibinfo {author} {\bibfnamefont
  {G.}~\bibnamefont {Leménager}}, \bibinfo {author} {\bibfnamefont
  {R.}~\bibnamefont {Houdré}}, \bibinfo {author} {\bibfnamefont
  {E}~\bibnamefont {Giacobino}}, \bibinfo {author} {\bibfnamefont
  {C.}~\bibnamefont {Ciuti}}, \ and\ \bibinfo {author} {\bibfnamefont
  {A.}~\bibnamefont {Bramati}},\ }\bibfield  {title} {\enquote {\bibinfo
  {title} {Polariton superfluids reveal quantum hydrodynamic solitons},}\
  }\href {\doibase 10.1126/science.1202307} {\bibfield  {journal} {\bibinfo
  {journal} {Science}\ }\textbf {\bibinfo {volume} {332}},\ \bibinfo {pages}
  {1167--1170} (\bibinfo {year} {2011})}\BibitemShut {NoStop}%
\bibitem [{\citenamefont {Carusotto}\ and\ \citenamefont
  {Ciuti}(2013)}]{RevModPhys.85.299}%
  \BibitemOpen
  \bibfield  {author} {\bibinfo {author} {\bibfnamefont {I.}~\bibnamefont
  {Carusotto}}\ and\ \bibinfo {author} {\bibfnamefont {C.}~\bibnamefont
  {Ciuti}},\ }\bibfield  {title} {\enquote {\bibinfo {title} {Quantum fluids of
  light},}\ }\href {\doibase 10.1103/RevModPhys.85.299} {\bibfield  {journal}
  {\bibinfo  {journal} {Rev. Mod. Phys.}\ }\textbf {\bibinfo {volume} {85}},\
  \bibinfo {pages} {299} (\bibinfo {year} {2013})}\BibitemShut {NoStop}%
\bibitem [{\citenamefont {Ozawa}\ \emph {et~al.}(2018)\citenamefont {Ozawa},
  \citenamefont {Price}, \citenamefont {Amo}, \citenamefont {Goldman},
  \citenamefont {Hafezi}, \citenamefont {Lu}, \citenamefont {Rechtsman},
  \citenamefont {Schuster}, \citenamefont {Simon}, \citenamefont {Zilberberg},\
  and\ \citenamefont {Carusotto}}]{1802.04173}%
  \BibitemOpen
  \bibfield  {author} {\bibinfo {author} {\bibfnamefont {T.}~\bibnamefont
  {Ozawa}}, \bibinfo {author} {\bibfnamefont {H.~M.}\ \bibnamefont {Price}},
  \bibinfo {author} {\bibfnamefont {A.}~\bibnamefont {Amo}}, \bibinfo {author}
  {\bibfnamefont {N.}~\bibnamefont {Goldman}}, \bibinfo {author} {\bibfnamefont
  {M.}~\bibnamefont {Hafezi}}, \bibinfo {author} {\bibfnamefont
  {L.}~\bibnamefont {Lu}}, \bibinfo {author} {\bibfnamefont {M.}~\bibnamefont
  {Rechtsman}}, \bibinfo {author} {\bibfnamefont {D.}~\bibnamefont {Schuster}},
  \bibinfo {author} {\bibfnamefont {J.}~\bibnamefont {Simon}}, \bibinfo
  {author} {\bibfnamefont {O.}~\bibnamefont {Zilberberg}}, \ and\ \bibinfo
  {author} {\bibfnamefont {I.}~\bibnamefont {Carusotto}},\ }\bibfield  {title}
  {\enquote {\bibinfo {title} {Topological photonics},}\ }\href@noop {} {\
  (\bibinfo {year} {2018})},\ \Eprint {http://arxiv.org/abs/1802.04173}
  {arXiv:1802.04173} \BibitemShut {NoStop}%
\bibitem [{\citenamefont {Dini}\ \emph {et~al.}(2003)\citenamefont {Dini},
  \citenamefont {K\"ohler}, \citenamefont {Tredicucci}, \citenamefont
  {Biasiol},\ and\ \citenamefont {Sorba}}]{PhysRevLett.90.116401}%
  \BibitemOpen
  \bibfield  {author} {\bibinfo {author} {\bibfnamefont {D.}~\bibnamefont
  {Dini}}, \bibinfo {author} {\bibfnamefont {R.}~\bibnamefont {K\"ohler}},
  \bibinfo {author} {\bibfnamefont {A.}~\bibnamefont {Tredicucci}}, \bibinfo
  {author} {\bibfnamefont {G.}~\bibnamefont {Biasiol}}, \ and\ \bibinfo
  {author} {\bibfnamefont {L.}~\bibnamefont {Sorba}},\ }\bibfield  {title}
  {\enquote {\bibinfo {title} {Microcavity polariton splitting of intersubband
  transitions},}\ }\href {\doibase 10.1103/PhysRevLett.90.116401} {\bibfield
  {journal} {\bibinfo  {journal} {Phys. Rev. Lett.}\ }\textbf {\bibinfo
  {volume} {90}},\ \bibinfo {pages} {116401} (\bibinfo {year}
  {2003})}\BibitemShut {NoStop}%
\bibitem [{\citenamefont {Dupont}\ \emph {et~al.}(2003)\citenamefont {Dupont},
  \citenamefont {Liu}, \citenamefont {SpringThorpe}, \citenamefont {Lai},\ and\
  \citenamefont {Extavour}}]{PhysRevB.68.245320}%
  \BibitemOpen
  \bibfield  {author} {\bibinfo {author} {\bibfnamefont {E.}~\bibnamefont
  {Dupont}}, \bibinfo {author} {\bibfnamefont {H.~C.}\ \bibnamefont {Liu}},
  \bibinfo {author} {\bibfnamefont {A.~J.}\ \bibnamefont {SpringThorpe}},
  \bibinfo {author} {\bibfnamefont {W.}~\bibnamefont {Lai}}, \ and\ \bibinfo
  {author} {\bibfnamefont {M.}~\bibnamefont {Extavour}},\ }\bibfield  {title}
  {\enquote {\bibinfo {title} {Vacuum-field {Rabi} splitting in quantum-well
  infrared photodetectors},}\ }\href {\doibase 10.1103/PhysRevB.68.245320}
  {\bibfield  {journal} {\bibinfo  {journal} {Phys. Rev. B}\ }\textbf {\bibinfo
  {volume} {68}},\ \bibinfo {pages} {245320} (\bibinfo {year}
  {2003})}\BibitemShut {NoStop}%
\bibitem [{\citenamefont {Ciuti}\ \emph {et~al.}(2005)\citenamefont {Ciuti},
  \citenamefont {Bastard},\ and\ \citenamefont {Carusotto}}]{ciuti2005quantum}%
  \BibitemOpen
  \bibfield  {author} {\bibinfo {author} {\bibfnamefont {C.}~\bibnamefont
  {Ciuti}}, \bibinfo {author} {\bibfnamefont {G.}~\bibnamefont {Bastard}}, \
  and\ \bibinfo {author} {\bibfnamefont {I.}~\bibnamefont {Carusotto}},\
  }\bibfield  {title} {\enquote {\bibinfo {title} {Quantum vacuum properties of
  the intersubband cavity polariton field},}\ }\href@noop {} {\bibfield
  {journal} {\bibinfo  {journal} {Physical Review B}\ }\textbf {\bibinfo
  {volume} {72}},\ \bibinfo {pages} {115303} (\bibinfo {year}
  {2005})}\BibitemShut {NoStop}%
\bibitem [{\citenamefont {Todorov}\ and\ \citenamefont
  {Sirtori}(2012)}]{PhysRevB.85.045304}%
  \BibitemOpen
  \bibfield  {author} {\bibinfo {author} {\bibfnamefont {Y.}~\bibnamefont
  {Todorov}}\ and\ \bibinfo {author} {\bibfnamefont {C.}~\bibnamefont
  {Sirtori}},\ }\bibfield  {title} {\enquote {\bibinfo {title} {Intersubband
  polaritons in the electrical dipole gauge},}\ }\href {\doibase
  10.1103/PhysRevB.85.045304} {\bibfield  {journal} {\bibinfo  {journal} {Phys.
  Rev. B}\ }\textbf {\bibinfo {volume} {85}},\ \bibinfo {pages} {045304}
  (\bibinfo {year} {2012})}\BibitemShut {NoStop}%
\bibitem [{\citenamefont {Anappara}\ \emph {et~al.}(2009)\citenamefont
  {Anappara}, \citenamefont {De~Liberato}, \citenamefont {Tredicucci},
  \citenamefont {Ciuti}, \citenamefont {Biasiol}, \citenamefont {Sorba},\ and\
  \citenamefont {Beltram}}]{PhysRevB.79.201303}%
  \BibitemOpen
  \bibfield  {author} {\bibinfo {author} {\bibfnamefont {A.~A.}\ \bibnamefont
  {Anappara}}, \bibinfo {author} {\bibfnamefont {S.}~\bibnamefont
  {De~Liberato}}, \bibinfo {author} {\bibfnamefont {A.}~\bibnamefont
  {Tredicucci}}, \bibinfo {author} {\bibfnamefont {C.}~\bibnamefont {Ciuti}},
  \bibinfo {author} {\bibfnamefont {G.}~\bibnamefont {Biasiol}}, \bibinfo
  {author} {\bibfnamefont {L.}~\bibnamefont {Sorba}}, \ and\ \bibinfo {author}
  {\bibfnamefont {F.}~\bibnamefont {Beltram}},\ }\bibfield  {title} {\enquote
  {\bibinfo {title} {Signatures of the ultrastrong light-matter coupling
  regime},}\ }\href {\doibase 10.1103/PhysRevB.79.201303} {\bibfield  {journal}
  {\bibinfo  {journal} {Phys. Rev. B}\ }\textbf {\bibinfo {volume} {79}},\
  \bibinfo {pages} {201303} (\bibinfo {year} {2009})}\BibitemShut {NoStop}%
\bibitem [{\citenamefont {Todorov}\ \emph {et~al.}(2010)\citenamefont
  {Todorov}, \citenamefont {Andrews}, \citenamefont {Colombelli}, \citenamefont
  {De~Liberato}, \citenamefont {Ciuti}, \citenamefont {Klang}, \citenamefont
  {Strasser},\ and\ \citenamefont {Sirtori}}]{PhysRevLett.105.196402}%
  \BibitemOpen
  \bibfield  {author} {\bibinfo {author} {\bibfnamefont {Y.}~\bibnamefont
  {Todorov}}, \bibinfo {author} {\bibfnamefont {A.~M.}\ \bibnamefont
  {Andrews}}, \bibinfo {author} {\bibfnamefont {R.}~\bibnamefont {Colombelli}},
  \bibinfo {author} {\bibfnamefont {S.}~\bibnamefont {De~Liberato}}, \bibinfo
  {author} {\bibfnamefont {C.}~\bibnamefont {Ciuti}}, \bibinfo {author}
  {\bibfnamefont {P.}~\bibnamefont {Klang}}, \bibinfo {author} {\bibfnamefont
  {G.}~\bibnamefont {Strasser}}, \ and\ \bibinfo {author} {\bibfnamefont
  {C.}~\bibnamefont {Sirtori}},\ }\bibfield  {title} {\enquote {\bibinfo
  {title} {Ultrastrong light-matter coupling regime with polariton dots},}\
  }\href {\doibase 10.1103/PhysRevLett.105.196402} {\bibfield  {journal}
  {\bibinfo  {journal} {Phys. Rev. Lett.}\ }\textbf {\bibinfo {volume} {105}},\
  \bibinfo {pages} {196402} (\bibinfo {year} {2010})}\BibitemShut {NoStop}%
\bibitem [{\citenamefont {Geiser}\ \emph {et~al.}(2012)\citenamefont {Geiser},
  \citenamefont {Castellano}, \citenamefont {Scalari}, \citenamefont {Beck},
  \citenamefont {Nevou},\ and\ \citenamefont {Faist}}]{PhysRevLett.108.106402}%
  \BibitemOpen
  \bibfield  {author} {\bibinfo {author} {\bibfnamefont {M.}~\bibnamefont
  {Geiser}}, \bibinfo {author} {\bibfnamefont {F.}~\bibnamefont {Castellano}},
  \bibinfo {author} {\bibfnamefont {G.}~\bibnamefont {Scalari}}, \bibinfo
  {author} {\bibfnamefont {M.}~\bibnamefont {Beck}}, \bibinfo {author}
  {\bibfnamefont {L.}~\bibnamefont {Nevou}}, \ and\ \bibinfo {author}
  {\bibfnamefont {J.}~\bibnamefont {Faist}},\ }\bibfield  {title} {\enquote
  {\bibinfo {title} {Ultrastrong coupling regime and plasmon polaritons in
  parabolic semiconductor quantum wells},}\ }\href {\doibase
  10.1103/PhysRevLett.108.106402} {\bibfield  {journal} {\bibinfo  {journal}
  {Phys. Rev. Lett.}\ }\textbf {\bibinfo {volume} {108}},\ \bibinfo {pages}
  {106402} (\bibinfo {year} {2012})}\BibitemShut {NoStop}%
\bibitem [{\citenamefont {Todorov}\ \emph {et~al.}(2009)\citenamefont
  {Todorov}, \citenamefont {Andrews}, \citenamefont {Sagnes}, \citenamefont
  {Colombelli}, \citenamefont {Klang}, \citenamefont {Strasser},\ and\
  \citenamefont {Sirtori}}]{PhysRevLett.102.186402}%
  \BibitemOpen
  \bibfield  {author} {\bibinfo {author} {\bibfnamefont {Y.}~\bibnamefont
  {Todorov}}, \bibinfo {author} {\bibfnamefont {A.~M.}\ \bibnamefont
  {Andrews}}, \bibinfo {author} {\bibfnamefont {I.}~\bibnamefont {Sagnes}},
  \bibinfo {author} {\bibfnamefont {R.}~\bibnamefont {Colombelli}}, \bibinfo
  {author} {\bibfnamefont {P.}~\bibnamefont {Klang}}, \bibinfo {author}
  {\bibfnamefont {G.}~\bibnamefont {Strasser}}, \ and\ \bibinfo {author}
  {\bibfnamefont {C.}~\bibnamefont {Sirtori}},\ }\bibfield  {title} {\enquote
  {\bibinfo {title} {Strong light-matter coupling in subwavelength
  metal-dielectric microcavities at terahertz frequencies},}\ }\href {\doibase
  10.1103/PhysRevLett.102.186402} {\bibfield  {journal} {\bibinfo  {journal}
  {Phys. Rev. Lett.}\ }\textbf {\bibinfo {volume} {102}},\ \bibinfo {pages}
  {186402} (\bibinfo {year} {2009})}\BibitemShut {NoStop}%
\bibitem [{\citenamefont {Zanotto}\ \emph {et~al.}(2012)\citenamefont
  {Zanotto}, \citenamefont {Degl'Innocenti}, \citenamefont {Xu}, \citenamefont
  {Sorba}, \citenamefont {Tredicucci},\ and\ \citenamefont
  {Biasiol}}]{PhysRevB.86.201302}%
  \BibitemOpen
  \bibfield  {author} {\bibinfo {author} {\bibfnamefont {S.}~\bibnamefont
  {Zanotto}}, \bibinfo {author} {\bibfnamefont {R.}~\bibnamefont
  {Degl'Innocenti}}, \bibinfo {author} {\bibfnamefont {J.-H.}\ \bibnamefont
  {Xu}}, \bibinfo {author} {\bibfnamefont {L.}~\bibnamefont {Sorba}}, \bibinfo
  {author} {\bibfnamefont {A.}~\bibnamefont {Tredicucci}}, \ and\ \bibinfo
  {author} {\bibfnamefont {G.}~\bibnamefont {Biasiol}},\ }\bibfield  {title}
  {\enquote {\bibinfo {title} {Ultrafast optical bleaching of intersubband
  cavity polaritons},}\ }\href {\doibase 10.1103/PhysRevB.86.201302} {\bibfield
   {journal} {\bibinfo  {journal} {Phys. Rev. B}\ }\textbf {\bibinfo {volume}
  {86}},\ \bibinfo {pages} {201302} (\bibinfo {year} {2012})}\BibitemShut
  {NoStop}%
\bibitem [{\citenamefont {Zanotto}\ \emph {et~al.}(2015)\citenamefont
  {Zanotto}, \citenamefont {Bianco}, \citenamefont {Sorba}, \citenamefont
  {Biasiol},\ and\ \citenamefont {Tredicucci}}]{PhysRevB.91.085308}%
  \BibitemOpen
  \bibfield  {author} {\bibinfo {author} {\bibfnamefont {S.}~\bibnamefont
  {Zanotto}}, \bibinfo {author} {\bibfnamefont {F.}~\bibnamefont {Bianco}},
  \bibinfo {author} {\bibfnamefont {L.}~\bibnamefont {Sorba}}, \bibinfo
  {author} {\bibfnamefont {G.}~\bibnamefont {Biasiol}}, \ and\ \bibinfo
  {author} {\bibfnamefont {A.}~\bibnamefont {Tredicucci}},\ }\bibfield  {title}
  {\enquote {\bibinfo {title} {Saturation and bistability of defect-mode
  intersubband polaritons},}\ }\href {\doibase 10.1103/PhysRevB.91.085308}
  {\bibfield  {journal} {\bibinfo  {journal} {Phys. Rev. B}\ }\textbf {\bibinfo
  {volume} {91}},\ \bibinfo {pages} {085308} (\bibinfo {year}
  {2015})}\BibitemShut {NoStop}%
\bibitem [{\citenamefont {De~Liberato}\ and\ \citenamefont
  {Ciuti}(2009)}]{PhysRevLett.102.136403}%
  \BibitemOpen
  \bibfield  {author} {\bibinfo {author} {\bibfnamefont {S.}~\bibnamefont
  {De~Liberato}}\ and\ \bibinfo {author} {\bibfnamefont {C.}~\bibnamefont
  {Ciuti}},\ }\bibfield  {title} {\enquote {\bibinfo {title} {Stimulated
  scattering and lasing of intersubband cavity polaritons},}\ }\href {\doibase
  10.1103/PhysRevLett.102.136403} {\bibfield  {journal} {\bibinfo  {journal}
  {Phys. Rev. Lett.}\ }\textbf {\bibinfo {volume} {102}},\ \bibinfo {pages}
  {136403} (\bibinfo {year} {2009})}\BibitemShut {NoStop}%
\bibitem [{\citenamefont {Nguyen-th\^e}\ \emph {et~al.}(2013)\citenamefont
  {Nguyen-th\^e}, \citenamefont {De~Liberato}, \citenamefont {Bamba},\ and\
  \citenamefont {Ciuti}}]{PhysRevB.87.235322}%
  \BibitemOpen
  \bibfield  {author} {\bibinfo {author} {\bibfnamefont {L.}~\bibnamefont
  {Nguyen-th\^e}}, \bibinfo {author} {\bibfnamefont {S.}~\bibnamefont
  {De~Liberato}}, \bibinfo {author} {\bibfnamefont {M.}~\bibnamefont {Bamba}},
  \ and\ \bibinfo {author} {\bibfnamefont {C.}~\bibnamefont {Ciuti}},\
  }\bibfield  {title} {\enquote {\bibinfo {title} {Effective
  polariton-polariton interactions of cavity-embedded two-dimensional electron
  gases},}\ }\href {\doibase 10.1103/PhysRevB.87.235322} {\bibfield  {journal}
  {\bibinfo  {journal} {Phys. Rev. B}\ }\textbf {\bibinfo {volume} {87}},\
  \bibinfo {pages} {235322} (\bibinfo {year} {2013})}\BibitemShut {NoStop}%
\bibitem [{\citenamefont {Colombelli}\ and\ \citenamefont
  {Manceau}(2015)}]{PhysRevX.5.011031}%
  \BibitemOpen
  \bibfield  {author} {\bibinfo {author} {\bibfnamefont {R.}~\bibnamefont
  {Colombelli}}\ and\ \bibinfo {author} {\bibfnamefont {J.-M.}\ \bibnamefont
  {Manceau}},\ }\bibfield  {title} {\enquote {\bibinfo {title} {Perspectives
  for intersubband polariton lasers},}\ }\href {\doibase
  10.1103/PhysRevX.5.011031} {\bibfield  {journal} {\bibinfo  {journal} {Phys.
  Rev. X}\ }\textbf {\bibinfo {volume} {5}},\ \bibinfo {pages} {011031}
  (\bibinfo {year} {2015})}\BibitemShut {NoStop}%
\bibitem [{\citenamefont {De~Liberato}\ \emph {et~al.}(2013)\citenamefont
  {De~Liberato}, \citenamefont {Ciuti},\ and\ \citenamefont
  {Phillips}}]{PhysRevB.87.241304}%
  \BibitemOpen
  \bibfield  {author} {\bibinfo {author} {\bibfnamefont {S.}~\bibnamefont
  {De~Liberato}}, \bibinfo {author} {\bibfnamefont {C.}~\bibnamefont {Ciuti}},
  \ and\ \bibinfo {author} {\bibfnamefont {C.~C.}\ \bibnamefont {Phillips}},\
  }\bibfield  {title} {\enquote {\bibinfo {title} {Terahertz lasing from
  intersubband polariton-polariton scattering in asymmetric quantum wells},}\
  }\href {\doibase 10.1103/PhysRevB.87.241304} {\bibfield  {journal} {\bibinfo
  {journal} {Phys. Rev. B}\ }\textbf {\bibinfo {volume} {87}},\ \bibinfo
  {pages} {241304} (\bibinfo {year} {2013})}\BibitemShut {NoStop}%
\bibitem [{\citenamefont {Hugi}\ \emph {et~al.}(2012)\citenamefont {Hugi},
  \citenamefont {Villares}, \citenamefont {Blaser}, \citenamefont {Liu},\ and\
  \citenamefont {Faist}}]{Nature.492.229}%
  \BibitemOpen
  \bibfield  {author} {\bibinfo {author} {\bibfnamefont {A.}~\bibnamefont
  {Hugi}}, \bibinfo {author} {\bibfnamefont {G.}~\bibnamefont {Villares}},
  \bibinfo {author} {\bibfnamefont {S.}~\bibnamefont {Blaser}}, \bibinfo
  {author} {\bibfnamefont {H.~C.}\ \bibnamefont {Liu}}, \ and\ \bibinfo
  {author} {\bibfnamefont {J.}~\bibnamefont {Faist}},\ }\bibfield  {title}
  {\enquote {\bibinfo {title} {Mid-infrared frequency comb based on a quantum
  cascade laser},}\ }\href {https://doi.org/10.1038/nature11620} {\bibfield
  {journal} {\bibinfo  {journal} {Nature}\ }\textbf {\bibinfo {volume} {492}},\
  \bibinfo {pages} {229} (\bibinfo {year} {2012})}\BibitemShut {NoStop}%
\bibitem [{\citenamefont {Vitiello}\ \emph {et~al.}(2015)\citenamefont
  {Vitiello}, \citenamefont {Scalari}, \citenamefont {Williams},\ and\
  \citenamefont {De~Natale}}]{OptExpress.23.5167}%
  \BibitemOpen
  \bibfield  {author} {\bibinfo {author} {\bibfnamefont {M.~S.}\ \bibnamefont
  {Vitiello}}, \bibinfo {author} {\bibfnamefont {G.}~\bibnamefont {Scalari}},
  \bibinfo {author} {\bibfnamefont {B.}~\bibnamefont {Williams}}, \ and\
  \bibinfo {author} {\bibfnamefont {P.}~\bibnamefont {De~Natale}},\ }\bibfield
  {title} {\enquote {\bibinfo {title} {Quantum cascade lasers: 20 years of
  challenges},}\ }\href {\doibase 10.1364/OE.23.005167} {\bibfield  {journal}
  {\bibinfo  {journal} {Opt. Express}\ }\textbf {\bibinfo {volume} {23}},\
  \bibinfo {pages} {5167--5182} (\bibinfo {year} {2015})}\BibitemShut {NoStop}%
\bibitem [{\citenamefont {Günter}\ \emph {et~al.}(2009)\citenamefont
  {Günter}, \citenamefont {Anappara}, \citenamefont {Hees}, \citenamefont
  {Sell}, \citenamefont {Biasiol}, \citenamefont {Sorba}, \citenamefont
  {De~Liberato}, \citenamefont {Ciuti}, \citenamefont {Tredicucci},
  \citenamefont {Leitenstorfer},\ and\ \citenamefont {Huber}}]{Nature.458.178}%
  \BibitemOpen
  \bibfield  {author} {\bibinfo {author} {\bibfnamefont {G.}~\bibnamefont
  {Günter}}, \bibinfo {author} {\bibfnamefont {A.~A.}\ \bibnamefont
  {Anappara}}, \bibinfo {author} {\bibfnamefont {J.}~\bibnamefont {Hees}},
  \bibinfo {author} {\bibfnamefont {A.}~\bibnamefont {Sell}}, \bibinfo {author}
  {\bibfnamefont {G.}~\bibnamefont {Biasiol}}, \bibinfo {author} {\bibfnamefont
  {L.}~\bibnamefont {Sorba}}, \bibinfo {author} {\bibfnamefont
  {S.}~\bibnamefont {De~Liberato}}, \bibinfo {author} {\bibfnamefont
  {C.}~\bibnamefont {Ciuti}}, \bibinfo {author} {\bibfnamefont
  {A.}~\bibnamefont {Tredicucci}}, \bibinfo {author} {\bibfnamefont
  {A.}~\bibnamefont {Leitenstorfer}}, \ and\ \bibinfo {author} {\bibfnamefont
  {R.}~\bibnamefont {Huber}},\ }\bibfield  {title} {\enquote {\bibinfo {title}
  {Sub-cycle switch-on of ultrastrong light–matter interaction},}\ }\href
  {https://doi.org/10.1038/nature07838} {\bibfield  {journal} {\bibinfo
  {journal} {Nature}\ }\textbf {\bibinfo {volume} {458}},\ \bibinfo {pages}
  {178} (\bibinfo {year} {2009})}\BibitemShut {NoStop}%
\bibitem [{\citenamefont {Leymann}\ and\ \citenamefont
  {Carusotto}(2019)}]{Leymann}%
  \BibitemOpen
  \bibfield  {author} {\bibinfo {author} {\bibfnamefont {H.~A.~M.}\
  \bibnamefont {Leymann}}\ and\ \bibinfo {author} {\bibfnamefont
  {I.}~\bibnamefont {Carusotto}},\ }\bibfield  {title} {\enquote {\bibinfo
  {title} {Nonlinear response of intersubband transitions in semiconductor
  quantum wells},}\ }\href@noop {} {\bibfield  {journal} {\bibinfo  {journal}
  {In preparation}\ } (\bibinfo {year} {2019})}\BibitemShut {NoStop}%
\bibitem [{\citenamefont {Manceau}\ \emph {et~al.}(2014)\citenamefont
  {Manceau}, \citenamefont {Zanotto}, \citenamefont {Ongarello}, \citenamefont
  {Sorba}, \citenamefont {Tredicucci}, \citenamefont {Biasiol},\ and\
  \citenamefont {Colombelli}}]{JMM2014}%
  \BibitemOpen
  \bibfield  {author} {\bibinfo {author} {\bibfnamefont {J.-M.}\ \bibnamefont
  {Manceau}}, \bibinfo {author} {\bibfnamefont {S.}~\bibnamefont {Zanotto}},
  \bibinfo {author} {\bibfnamefont {T.}~\bibnamefont {Ongarello}}, \bibinfo
  {author} {\bibfnamefont {L.}~\bibnamefont {Sorba}}, \bibinfo {author}
  {\bibfnamefont {A.}~\bibnamefont {Tredicucci}}, \bibinfo {author}
  {\bibfnamefont {G.}~\bibnamefont {Biasiol}}, \ and\ \bibinfo {author}
  {\bibfnamefont {R.}~\bibnamefont {Colombelli}},\ }\bibfield  {title}
  {\enquote {\bibinfo {title} {Mid-infrared intersubband polaritons in
  dispersive metal-insulator-metal resonators},}\ }\href {\doibase
  10.1063/1.4893730} {\bibfield  {journal} {\bibinfo  {journal} {Appl. Phys.
  Lett.}\ }\textbf {\bibinfo {volume} {105}},\ \bibinfo {pages} {081105}
  (\bibinfo {year} {2014})}\BibitemShut {NoStop}%
\bibitem [{\citenamefont {Manceau}\ \emph {et~al.}(2017)\citenamefont
  {Manceau}, \citenamefont {Biasiol}, \citenamefont {Tran}, \citenamefont
  {Carusotto},\ and\ \citenamefont {Colombelli}}]{PhysRevB.96.235301}%
  \BibitemOpen
  \bibfield  {author} {\bibinfo {author} {\bibfnamefont {J-M.}\ \bibnamefont
  {Manceau}}, \bibinfo {author} {\bibfnamefont {G.}~\bibnamefont {Biasiol}},
  \bibinfo {author} {\bibfnamefont {N.~L.}\ \bibnamefont {Tran}}, \bibinfo
  {author} {\bibfnamefont {I.}~\bibnamefont {Carusotto}}, \ and\ \bibinfo
  {author} {\bibfnamefont {R.}~\bibnamefont {Colombelli}},\ }\bibfield  {title}
  {\enquote {\bibinfo {title} {Immunity of intersubband polaritons to
  inhomogeneous broadening},}\ }\href {\doibase 10.1103/PhysRevB.96.235301}
  {\bibfield  {journal} {\bibinfo  {journal} {Phys. Rev. B}\ }\textbf {\bibinfo
  {volume} {96}},\ \bibinfo {pages} {235301} (\bibinfo {year}
  {2017})}\BibitemShut {NoStop}%
\bibitem [{\citenamefont {Savvidis}\ \emph {et~al.}(2000)\citenamefont
  {Savvidis}, \citenamefont {Baumberg}, \citenamefont {Stevenson},
  \citenamefont {Skolnick}, \citenamefont {Whittaker},\ and\ \citenamefont
  {Roberts}}]{PhysRevLett.84.1547}%
  \BibitemOpen
  \bibfield  {author} {\bibinfo {author} {\bibfnamefont {P.~G.}\ \bibnamefont
  {Savvidis}}, \bibinfo {author} {\bibfnamefont {J.~J.}\ \bibnamefont
  {Baumberg}}, \bibinfo {author} {\bibfnamefont {R.~M.}\ \bibnamefont
  {Stevenson}}, \bibinfo {author} {\bibfnamefont {M.~S.}\ \bibnamefont
  {Skolnick}}, \bibinfo {author} {\bibfnamefont {D.~M.}\ \bibnamefont
  {Whittaker}}, \ and\ \bibinfo {author} {\bibfnamefont {J.~S.}\ \bibnamefont
  {Roberts}},\ }\bibfield  {title} {\enquote {\bibinfo {title} {Angle-resonant
  stimulated polariton amplifier},}\ }\href {\doibase
  10.1103/PhysRevLett.84.1547} {\bibfield  {journal} {\bibinfo  {journal}
  {Phys. Rev. Lett.}\ }\textbf {\bibinfo {volume} {84}},\ \bibinfo {pages}
  {1547--1550} (\bibinfo {year} {2000})}\BibitemShut {NoStop}%
\bibitem [{\citenamefont {Baumberg}\ \emph {et~al.}(2000)\citenamefont
  {Baumberg}, \citenamefont {Savvidis}, \citenamefont {Stevenson},
  \citenamefont {Tartakovskii}, \citenamefont {Skolnick}, \citenamefont
  {Whittaker},\ and\ \citenamefont {Roberts}}]{PhysRevB.62.R16247}%
  \BibitemOpen
  \bibfield  {author} {\bibinfo {author} {\bibfnamefont {J.~J.}\ \bibnamefont
  {Baumberg}}, \bibinfo {author} {\bibfnamefont {P.~G.}\ \bibnamefont
  {Savvidis}}, \bibinfo {author} {\bibfnamefont {R.~M.}\ \bibnamefont
  {Stevenson}}, \bibinfo {author} {\bibfnamefont {A.~I.}\ \bibnamefont
  {Tartakovskii}}, \bibinfo {author} {\bibfnamefont {M.~S.}\ \bibnamefont
  {Skolnick}}, \bibinfo {author} {\bibfnamefont {D.~M.}\ \bibnamefont
  {Whittaker}}, \ and\ \bibinfo {author} {\bibfnamefont {J.~S.}\ \bibnamefont
  {Roberts}},\ }\bibfield  {title} {\enquote {\bibinfo {title} {Parametric
  oscillation in a vertical microcavity: A polariton condensate or
  micro-optical parametric oscillation},}\ }\href {\doibase
  10.1103/PhysRevB.62.R16247} {\bibfield  {journal} {\bibinfo  {journal} {Phys.
  Rev. B}\ }\textbf {\bibinfo {volume} {62}},\ \bibinfo {pages} {R16247}
  (\bibinfo {year} {2000})}\BibitemShut {NoStop}%
\bibitem [{\citenamefont {Keeling}\ \emph {et~al.}(2007)\citenamefont
  {Keeling}, \citenamefont {Marchetti}, \citenamefont {Szyma{\'{n}}ska},\ and\
  \citenamefont {Littlewood}}]{SemicondSciTechnol.22.R1}%
  \BibitemOpen
  \bibfield  {author} {\bibinfo {author} {\bibfnamefont {J.}~\bibnamefont
  {Keeling}}, \bibinfo {author} {\bibfnamefont {F.~M.}\ \bibnamefont
  {Marchetti}}, \bibinfo {author} {\bibfnamefont {M.~H.}\ \bibnamefont
  {Szyma{\'{n}}ska}}, \ and\ \bibinfo {author} {\bibfnamefont {P.~B.}\
  \bibnamefont {Littlewood}},\ }\bibfield  {title} {\enquote {\bibinfo {title}
  {Collective coherence in planar semiconductor microcavities},}\ }\href
  {\doibase 10.1088/0268-1242/22/5/r01} {\bibfield  {journal} {\bibinfo
  {journal} {Semicond. Sci. Technol.}\ }\textbf {\bibinfo {volume} {22}},\
  \bibinfo {pages} {R1} (\bibinfo {year} {2007})}\BibitemShut {NoStop}%
\bibitem [{\citenamefont {Breuer}\ and\ \citenamefont
  {Petruccione}(2002)}]{Petruccione}%
  \BibitemOpen
  \bibfield  {author} {\bibinfo {author} {\bibfnamefont {H.~P.}\ \bibnamefont
  {Breuer}}\ and\ \bibinfo {author} {\bibfnamefont {F.}~\bibnamefont
  {Petruccione}},\ }\href@noop {} {\emph {\bibinfo {title} {The theory of open
  quantum systems}}}\ (\bibinfo  {publisher} {Oxford University Press},\
  \bibinfo {year} {2002})\BibitemShut {NoStop}%
\bibitem [{\citenamefont {Cohen-Tannoudji}\ \emph {et~al.}(2004)\citenamefont
  {Cohen-Tannoudji}, \citenamefont {Dupont-Roc},\ and\ \citenamefont
  {Grynberg}}]{cohen1998}%
  \BibitemOpen
  \bibfield  {author} {\bibinfo {author} {\bibfnamefont {C.}~\bibnamefont
  {Cohen-Tannoudji}}, \bibinfo {author} {\bibfnamefont {J.}~\bibnamefont
  {Dupont-Roc}}, \ and\ \bibinfo {author} {\bibfnamefont {G.}~\bibnamefont
  {Grynberg}},\ }\href@noop {} {\emph {\bibinfo {title} {Atom-photon
  interactions: basic processes and applications}}}\ (\bibinfo  {publisher}
  {Wiley-VCH},\ \bibinfo {address} {Weinheim},\ \bibinfo {year}
  {2004})\BibitemShut {NoStop}%
\bibitem [{\citenamefont {Wu}\ and\ \citenamefont
  {Zhang}(1997)}]{ApplPhysLett.71.1285}%
  \BibitemOpen
  \bibfield  {author} {\bibinfo {author} {\bibfnamefont {Q.}~\bibnamefont
  {Wu}}\ and\ \bibinfo {author} {\bibfnamefont {X.-C.}\ \bibnamefont {Zhang}},\
  }\bibfield  {title} {\enquote {\bibinfo {title} {Free-space electro-optics
  sampling of mid-infrared pulses},}\ }\href {\doibase 10.1063/1.119873}
  {\bibfield  {journal} {\bibinfo  {journal} {Appl. Phys. Lett.}\ }\textbf
  {\bibinfo {volume} {71}},\ \bibinfo {pages} {1285} (\bibinfo {year}
  {1997})}\BibitemShut {NoStop}%
\bibitem [{\citenamefont {Huber}\ \emph {et~al.}(2000)\citenamefont {Huber},
  \citenamefont {Brodschelm}, \citenamefont {Tauser},\ and\ \citenamefont
  {Leitenstorfer}}]{ApplPhysLett.76.3191}%
  \BibitemOpen
  \bibfield  {author} {\bibinfo {author} {\bibfnamefont {R.}~\bibnamefont
  {Huber}}, \bibinfo {author} {\bibfnamefont {A.}~\bibnamefont {Brodschelm}},
  \bibinfo {author} {\bibfnamefont {F.}~\bibnamefont {Tauser}}, \ and\ \bibinfo
  {author} {\bibfnamefont {A.}~\bibnamefont {Leitenstorfer}},\ }\bibfield
  {title} {\enquote {\bibinfo {title} {Generation and field-resolved detection
  of femtosecond electromagnetic pulses tunable up to 41 {THz}},}\ }\href
  {\doibase 10.1063/1.126625} {\bibfield  {journal} {\bibinfo  {journal} {Appl.
  Phys. Lett.}\ }\textbf {\bibinfo {volume} {76}},\ \bibinfo {pages} {3191}
  (\bibinfo {year} {2000})}\BibitemShut {NoStop}%
\bibitem [{\citenamefont {Liu}\ \emph {et~al.}(2004)\citenamefont {Liu},
  \citenamefont {Xu},\ and\ \citenamefont {Zhang}}]{ApplPhysLett.85.863}%
  \BibitemOpen
  \bibfield  {author} {\bibinfo {author} {\bibfnamefont {K.}~\bibnamefont
  {Liu}}, \bibinfo {author} {\bibfnamefont {J.}~\bibnamefont {Xu}}, \ and\
  \bibinfo {author} {\bibfnamefont {X.-C.}\ \bibnamefont {Zhang}},\ }\bibfield
  {title} {\enquote {\bibinfo {title} {{GaSe} crystals for broadband terahertz
  wave detection},}\ }\href {\doibase 10.1063/1.1779959} {\bibfield  {journal}
  {\bibinfo  {journal} {Appl. Phys. Lett.}\ }\textbf {\bibinfo {volume} {85}},\
  \bibinfo {pages} {863} (\bibinfo {year} {2004})}\BibitemShut {NoStop}%
\bibitem [{\citenamefont {Kübler}\ \emph {et~al.}(2004)\citenamefont
  {Kübler}, \citenamefont {Huber}, \citenamefont {Tübel},\ and\ \citenamefont
  {Leitenstorfer}}]{ApplPhysLett.85.3360}%
  \BibitemOpen
  \bibfield  {author} {\bibinfo {author} {\bibfnamefont {C.}~\bibnamefont
  {Kübler}}, \bibinfo {author} {\bibfnamefont {R.}~\bibnamefont {Huber}},
  \bibinfo {author} {\bibfnamefont {S.}~\bibnamefont {Tübel}}, \ and\ \bibinfo
  {author} {\bibfnamefont {A.}~\bibnamefont {Leitenstorfer}},\ }\bibfield
  {title} {\enquote {\bibinfo {title} {Ultrabroadband detection of
  multi-terahertz field transients with {GaSe} electro-optic sensors:
  Approaching the near infrared},}\ }\href {\doibase 10.1063/1.1808232}
  {\bibfield  {journal} {\bibinfo  {journal} {Appl. Phys. Lett.}\ }\textbf
  {\bibinfo {volume} {85}},\ \bibinfo {pages} {3360} (\bibinfo {year}
  {2004})}\BibitemShut {NoStop}%
\bibitem [{\citenamefont {Porer}\ \emph {et~al.}(2014)\citenamefont {Porer},
  \citenamefont {M\'{e}nard},\ and\ \citenamefont {Huber}}]{OptLett.8.2435}%
  \BibitemOpen
  \bibfield  {author} {\bibinfo {author} {\bibfnamefont {M.}~\bibnamefont
  {Porer}}, \bibinfo {author} {\bibfnamefont {J.-M.}\ \bibnamefont
  {M\'{e}nard}}, \ and\ \bibinfo {author} {\bibfnamefont {R.}~\bibnamefont
  {Huber}},\ }\bibfield  {title} {\enquote {\bibinfo {title} {Shot noise
  reduced terahertz detection via spectrally postfiltered electro-optic
  sampling},}\ }\href {\doibase 10.1364/OL.39.002435} {\bibfield  {journal}
  {\bibinfo  {journal} {Opt. Lett.}\ }\textbf {\bibinfo {volume} {39}},\
  \bibinfo {pages} {2435} (\bibinfo {year} {2014})}\BibitemShut {NoStop}%
\bibitem [{\citenamefont {{Keiber}}\ \emph {et~al.}(2016)\citenamefont
  {{Keiber}}, \citenamefont {{Sederberg}}, \citenamefont {{Schwarz}},
  \citenamefont {{Trubetskov}}, \citenamefont {{Pervak}}, \citenamefont
  {{Krausz}},\ and\ \citenamefont {{Karpowicz}}}]{NaturePhotonics.10.159}%
  \BibitemOpen
  \bibfield  {author} {\bibinfo {author} {\bibfnamefont {S.}~\bibnamefont
  {{Keiber}}}, \bibinfo {author} {\bibfnamefont {S.}~\bibnamefont
  {{Sederberg}}}, \bibinfo {author} {\bibfnamefont {A.}~\bibnamefont
  {{Schwarz}}}, \bibinfo {author} {\bibfnamefont {M.}~\bibnamefont
  {{Trubetskov}}}, \bibinfo {author} {\bibfnamefont {V.}~\bibnamefont
  {{Pervak}}}, \bibinfo {author} {\bibfnamefont {F.}~\bibnamefont {{Krausz}}},
  \ and\ \bibinfo {author} {\bibfnamefont {N.}~\bibnamefont {{Karpowicz}}},\
  }\bibfield  {title} {\enquote {\bibinfo {title} {Electro-optic sampling of
  near-infrared waveforms},}\ }\href {\doibase 10.1038/nphoton.2015.269}
  {\bibfield  {journal} {\bibinfo  {journal} {Nature Photonics}\ }\textbf
  {\bibinfo {volume} {10}},\ \bibinfo {pages} {159} (\bibinfo {year}
  {2016})}\BibitemShut {NoStop}%
\bibitem [{\citenamefont {Knorr}\ \emph {et~al.}(2017)\citenamefont {Knorr},
  \citenamefont {Raab}, \citenamefont {Tauer}, \citenamefont {Merkl},
  \citenamefont {Peller}, \citenamefont {Wittmann}, \citenamefont {Riedle},
  \citenamefont {Lange},\ and\ \citenamefont {Huber}}]{OptLett.21.4367}%
  \BibitemOpen
  \bibfield  {author} {\bibinfo {author} {\bibfnamefont {M.}~\bibnamefont
  {Knorr}}, \bibinfo {author} {\bibfnamefont {J.}~\bibnamefont {Raab}},
  \bibinfo {author} {\bibfnamefont {M.}~\bibnamefont {Tauer}}, \bibinfo
  {author} {\bibfnamefont {P.}~\bibnamefont {Merkl}}, \bibinfo {author}
  {\bibfnamefont {D.}~\bibnamefont {Peller}}, \bibinfo {author} {\bibfnamefont
  {E.}~\bibnamefont {Wittmann}}, \bibinfo {author} {\bibfnamefont
  {E.}~\bibnamefont {Riedle}}, \bibinfo {author} {\bibfnamefont
  {C.}~\bibnamefont {Lange}}, \ and\ \bibinfo {author} {\bibfnamefont
  {R.}~\bibnamefont {Huber}},\ }\bibfield  {title} {\enquote {\bibinfo {title}
  {Phase-locked multi-terahertz electric fields exceeding 13 {MV/cm} at a 190
  {kHz} repetition rate},}\ }\href {\doibase 10.1364/OL.42.004367} {\bibfield
  {journal} {\bibinfo  {journal} {Opt. Lett.}\ }\textbf {\bibinfo {volume}
  {42}},\ \bibinfo {pages} {4367} (\bibinfo {year} {2017})}\BibitemShut
  {NoStop}%
\bibitem [{Note1()}]{Note1}%
  \BibitemOpen
  \bibinfo {note} {To obtain a better match with experimental data, we had to
  slightly modify the photon dispersion relation in the region of interest to
  \begin {equation} \omega _C(k) = |k|/n + \omega _0, \end {equation} with a
  small $\omega _0 = \protect \unit {20}{\protect \ensuremath {\protect \@text
  {\protect \milli \protect \ensuremath {\protect \@text {\protect
  \electronvolt \protect \xspace }}}}}$, so to include the consequences of,
  e.g., the leakage of electromagnetic radiation inside the metallic
  mirrors.}\BibitemShut {Stop}%
\bibitem [{Note2()}]{Note2}%
  \BibitemOpen
  \bibinfo {note} {Another pump-only strategy to estimate polariton
  interactions is of course based on the electro-optical sampling of the
  time-dependence of the emission frequency during ring-down oscillation after
  a strong pulsed pump.}\BibitemShut {Stop}%
\bibitem [{\citenamefont {Ciuti}\ and\ \citenamefont
  {Carusotto}(2006)}]{Ciuti:2006PRA}%
  \BibitemOpen
  \bibfield  {author} {\bibinfo {author} {\bibfnamefont {C.}~\bibnamefont
  {Ciuti}}\ and\ \bibinfo {author} {\bibfnamefont {I.}~\bibnamefont
  {Carusotto}},\ }\bibfield  {title} {\enquote {\bibinfo {title} {Input-output
  theory of cavities in the ultrastrong coupling regime: The case of
  time-independent cavity parameters},}\ }\href {\doibase
  10.1103/PhysRevA.74.033811} {\bibfield  {journal} {\bibinfo  {journal} {Phys.
  Rev. A}\ }\textbf {\bibinfo {volume} {74}},\ \bibinfo {pages} {033811}
  (\bibinfo {year} {2006})}\BibitemShut {NoStop}%
\end{thebibliography}%
\end{document}